\documentclass[aps,prb,10pt,twocolumn,floatfix,footinbib,superscriptaddress,longbibliography,nobibnotes]{revtex4-1}

\usepackage{amsmath}
\usepackage{amssymb}
\usepackage{mathtools}
\usepackage{wasysym}

\usepackage[T1]{fontenc}
\usepackage{amsfonts}
\usepackage{newtxtext}
\usepackage[varvw]{newtxmath}
\usepackage{dsfont}                 % double stroke font
\usepackage{bbold}                  % \mathbb{0}
\usepackage[normalem]{ulem}         % \xout

\usepackage{enumerate}
\usepackage[shortlabels]{enumitem}

\usepackage[dvipsnames,table]{xcolor}
\usepackage{graphicx}
\graphicspath{{figs/}} % Setting the graphics path

\usepackage{physics}   % physics notation
\usepackage{csquotes}  % quotations
\usepackage{comment}   % comment blocks of text
\usepackage{relsize}   % smaller font for panel labels a,b,c,...

\usepackage[caption=false]{subfig}
\definecolor{naturelink}{HTML}{025E8D}
\usepackage[
  colorlinks=true,
  linkcolor=naturelink,
  citecolor=naturelink,
  urlcolor=naturelink,
  filecolor=naturelink,
  bookmarksnumbered
]{hyperref}
\usepackage[capitalize]{cleveref}

\usepackage[globalcitecopy]{bibunits}
\defaultbibliography{ref}
\defaultbibliographystyle{apsrev4-1}

\usepackage{orcidlink}

\def\imi{\textrm{i}}
\def\imj{\textrm{j}}
\def\imk{\textrm{k}}

\def\bs#1{{\mathbf{#1}}}
\def\bfsf#1{\textsf{\textbf{\smaller #1}}}

\makeatletter
\newif\ifnatc@aftersection
\natc@aftersectionfalse

\newcommand{\sectitle}[1]{%
  \par\addvspace{\medskipamount}%
  \noindent\phantomsection
  {\large\bfseries #1\par}%
  \@afterindentfalse\@afterheading
  \global\natc@aftersectiontrue
  \nopagebreak
}

\newcommand{\subsectitle}[1]{%
  \par\addvspace{\ifnatc@aftersection 0pt\else \medskipamount\fi}%
  \noindent\phantomsection
  {\bfseries #1\par}%
  \@afterindentfalse\@afterheading
  \global\natc@aftersectionfalse
  \nopagebreak
}

\makeatother

\definecolor{edit}{rgb}{0.0,0.0,0.8}

\definecolor{remove}{rgb}{0.8,0.0,0.0}

 \definecolor{TB}{rgb}{0,0.52,0.42}
\definecolor{ZG}{rgb}{1,0.2,0.2}

\definecolor{LU}{rgb}{0.7, 0.3, 0.8}

\begin{document}
\begin{bibunit}

\title{Topological non-Abelian gauge structures in Cayley-Schreier lattices}

\author{Zoltán Guba\,\orcidlink{0000-0002-6130-1064}}
\affiliation{Department of Physics, University of Z\"urich, Winterthurerstrasse 190, 8057 Z\"urich, Switzerland}
\author{Robert-Jan Slager\,\orcidlink{0000-0001-9055-5218}}
\affiliation{Department of Physics and Astronomy, University of Manchester,
Oxford Road, Manchester M13 9PL, United Kingdom}
\author{Lavi K.~Upreti\,\orcidlink{0000-0002-1722-484X}}\email{lavi.upreti@uzh.ch}
\affiliation{Department of Physics, University of Z\"urich, Winterthurerstrasse 190, 8057 Z\"urich, Switzerland}
\author{Tom\'a\v{s} Bzdu\v{s}ek\,\orcidlink{0000-0001-6904-5264}}\email{tomas.bzdusek@uzh.ch}
\affiliation{Department of Physics, University of Z\"urich, Winterthurerstrasse 190, 8057 Z\"urich, Switzerland}

%%%%%%%%%%%%%%%%%%%
%%%  ABSTRACT   %%%
%%%%%%%%%%%%%%%%%%%

\begin{abstract}
\sectitle{Abstract}
Crystalline constructions known as Cayley-Schreier lattices have been suggested as a platform for realizing arbitrary gauge fields in synthetic crystals with real hopping amplitudes. Here, we reveal
that Cayley-Schreier lattices can naturally give rise to implementable lattice systems that incorporate non-Abelian gauge structures transforming into a space-group symmetry. We show that their 
symmetry sectors can be interpreted as blocks of pseudospin models—some of which correspond to true spinors—that can realize a wealth of different topological invariants in a single setup. We underpin these general results with concrete models and illustrate 
how they can be implemented in technologically available experimental platforms. Our work sets the stage for a systematic investigation of topological insulators and metals with non-Abelian gauge structures.
\end{abstract}

\maketitle
%TC:endignore

%%%%%%%%%%%%%%%%%%%%%
%%% Introduction  %%%
%%%%%%%%%%%%%%%%%%%%%

\sectitle{Introduction}
Although a fairly uniform understanding of conventional topological band theory~\cite{Rmp1, Rmp2, Ryu:2010, kruthoff2017, Po2017, Bradlyn2017} has emerged over the past decade, recent results have opened further active research directions within this field. 
These, in particular, include multi-gap topological phases~\cite{Wu:2018,Bouhon:2020,Bouhon2020,Tiwari:2020,Jiang:2021,slager2024floquet,Guo2021Dexp} in which invariants 
emerge from simultaneous involvement of multiple bands, typified by systems with ${\cal I=PT}$ symmetry (product of parity and time reversal) with ${\cal I}^2=1$, and systems with projectively represented space-group symmetry~\cite{Zhao_2020,Chen:2022,Chen2023,Zhao:2021,Zhao_2024,Chen:2025}.
The latter direction is well exemplified by spinless systems whose
spatial-temporal symmetry obeys 
${\cal I}^2=-1$~\cite{Zhao_2020,Zhao:2021} due to modifying the symmetry operators with gauge transformations.
This line of investigations is particularly compelling, as it combines the concepts of gauge theory~\cite{Kogut1979}---a topic currently heavily explored in quantum simulators~\cite{Banuls2020}---with symmetry representations of band theory. 
In particular, when the symmetry group is 
represented projectively, meaning that the representation of translations, reflections, or other symmetries admits an additional phase factor, these phases can add up to a finite value along a closed contour, or, in other words, to a gauge flux~\cite{Zhao_2020,Zhao:2021,Chen:2022,Chen2023,Zhao_2024,Chen:2025}.
This shows that altered symmetry can underlie topological band classification, which has been experimentally verified~\cite{Xue:2022,Li:2022,Li2023acoustic,Pu2023klein,Lai2024phononic,ZHU2024,teo2025observationembeddedtopologytrivial} in various realizations. 
Nonetheless, this progress has, by and large, been limited to the $Z_2$ gauge group, thus posing the question of whether larger gauge groups---possibly non-Abelian ones---may enable previously inaccessible phenomena.

In a separate context, an alternative route of incorporating non-commutativity of translations and effective synthetic gauge structures has been explored through constructions called Cayley-Schreier lattices (CSLs)~\cite{Lux2024,Marciani_2025}.
Such lattices, in essence, proceed by generalizing the ``on-site'' symmetry group, producing a Hamiltonian whose translations allow for more generic structures. 
However, here also the focus has been on Abelian $Z_n$ cases. 
Given the freedom in defining the on-site symmetry group, this provokes the central question of whether non-Abelian structures can similarly be introduced while enabling new physical manifestations. 

Building on these developments, we here affirmatively answer the central question presented above and formulate a broader class of non-Abelian CSLs. As a main result, we find that CSLs provide a platform in which non-Abelian gauge structures naturally emerge from space group symmetries.
As a next step, we show that the dynamics on a CSL separates into decoupled blocks, each corresponding to a symmetry sector of the gauge group. 
The different symmetry sectors can be interpreted as pseudospin-$N$ models---some of which correspond to true $(2N\,{+}\,1)$-component spinors---where $N$ is specified by the dimension of the block.
This moreover allows for a route towards topology. 
Indeed, we motivate that for the two-component spinor (that is, spin-$\frac{1}{2}$ fermions exemplified by electrons) all known types of band topology can be induced within this single framework.
We finally illustrate this general theory with concrete examples in one and two spatial dimensions. 
These models can directly serve as input for meta-material realizations.
To that end, we also provide specific pathways for experimental validation in electric-circuit networks~\cite{Ningyuan:2015,Lee:2018,Imhof:2018,Helbig:2019,Hofmann:2020,Lenggenhager:2021,Kotwal:2021,Yatsugi:2022,Chen:2023b} with concrete proposals.

\begin{figure*}[t]
  \centering 
  \includegraphics[width=0.99\linewidth]{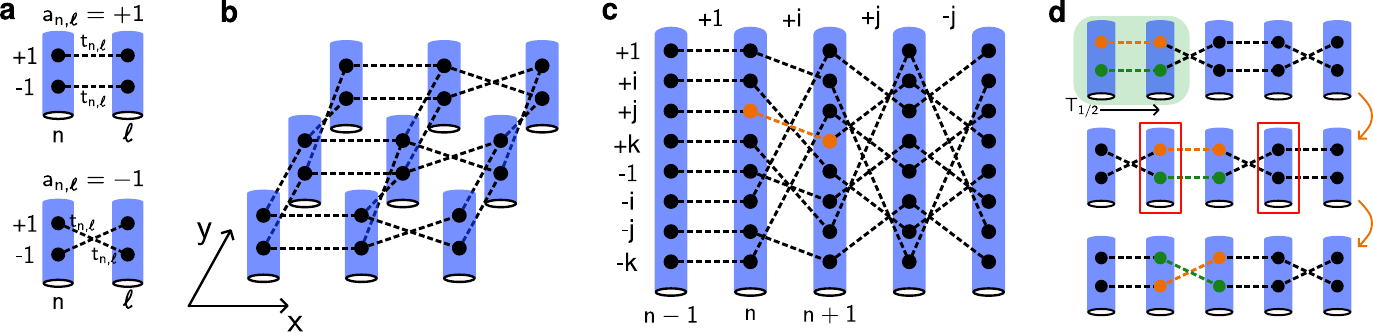}
  \caption{
  \textbf{Synthetic gauge fields in Cayley-Schreier lattices.}
  \bfsf{a}~Pair of connected pillars of a Cayley-Schreier lattice (CSL) with gauge group $G= Z_2$. 
  Each pillar represents a lattice site (indices $n$ and $\ell$) and contains two orbitals (one for each group element $\pm 1\in Z_2$). 
  The bonds between the constituent orbitals are `horizontal' (top) or `swapped' (bottom), depending on the value of the connection $a_{n,\ell} = \pm 1 \in Z_2$.
  \bfsf{b}~Illustration of a CSL with $G=Z_2$ over two-dimensional lattice.
  One of the two choices of connection from panel \bfsf{a} is made for the bonds between any pair of connected pillars. 
  \bfsf{c}~Schematics of a CSL over a one-dimensional chain with gauge group $G=Q_8$. 
  There are $\abs{Q_8}=8$ choices of connection possible on each bond, of which four (namely $a_{n+1,n}\in\{+1,+\imi,+\imj,-\imj\}$) are illustrated.
  The components highlighted with orange indicate that the orbital $g_n=+\imj$ at site $n$ is coupled by connection $a_{n+1,n} = +\imi$ to the orbital $g_{n+1} = a_{n+1,n}\cdot g_n = (+\imi) \cdot (+\imj) = +\imk$ at site $n+1$.
  \bfsf{d}~Alternating arrangement of trivial and non-trivial connections in a one-dimensional CSL with gauge group $Z_2$ (top). 
  Translation by one site flips the connection pattern (middle).
  Applying gauge transformation $\chi_i = -1$ to every second site (red rectangles) restores the original connection (bottom).
  The coloring of orbitals and hoppings within the two selected sites serve as a guide to follow the application of the translation and of the subsequent gauge transformation. 
  In panels \bfsf{b}--\bfsf{d},   dashed lines represent couplings of the same strength $t$.
  }
  \label{fig:CSL+GT}
\end{figure*}

%%%%%%%%%%%%%%%%
%%% Results  %%%
%%%%%%%%%%%%%%%%

\sectitle{Results}
\subsectitle{Cayley-Schreier lattices}
We now review the concept of a Cayley-Schreier lattice (CSL). 
The key insight here is that the conventional sites are replaced with ``pillars'' of internal degrees of freedom that transform under a finite group $G$. 
While our results can be directly generalized, we focus in particular on the case where this symmetry group entails the quaternion group 
\begin{equation}
\label{eqn:quaternions}
Q_8 = \{\pm 1,\pm \imi,\pm\imj,\pm\imk\},    
\end{equation} 
where $\imi$, $\imj$, and $\imk=\imi\cdot\imj$ are pairwise anticommuting units squaring to $\imi^2 = \imj^2 =\imk^2 = -1$.
We also adopt the group $Z_2 = \{\pm 1\}$ to clarify the building blocks of CSLs.

In the CSL construction, illustrated in Fig.~\ref{fig:CSL+GT}\bfsf{a}, each pillar consists of $|G|$ orbitals (depicted as black dots), with each orbital labeled by a distinct group element $g\,{\in}\, G$. Here $|G|$ is the number of elements in the group $G$. 
For $G= Z_2$, as an elementary example, each pillar has two orbitals, whereas for $G=Q_8$, each pillar hosts eight orbitals corresponding to the eight elements in Eq.~(\ref{eqn:quaternions}). A hopping Hamiltonian is then formed by connecting the orbitals of different pillars, and thereby generating the bonds of the lattice with a given space group~($\mathsf{SG}$).

Most importantly, these connections are chosen so that they respect the group multiplication in the following sense: 
we assign a group element $a_{n,\ell}\,{\in}\, G$ to the bond from site $\ell$ to site $n$ 
and ensure that orbital $g_{\ell}$ at $\ell$ is connected to orbital $a_{n,\ell} \cdot g_{\ell}$ at $n$ (the dot stands for the binary composition in $G$).
For the $Z_2$ example, this yields two types of bonds: `horizontal' for $a_{n,\ell}=+1$ and `swapped' for $a_{n,\ell}=-1$, both featured as dashed lines in Fig.~\ref{fig:CSL+GT}\bfsf{a} and \bfsf{b}.
Several allowed connections $a_{n,\ell}$ between the pillars for the case $G=Q_8$ are illustrated in Fig.~\ref{fig:CSL+GT}\bfsf{c}. 
This construction ensures that between any two connected pillars $n$ and $\ell$, there are $|G|$ hopping terms; namely, one for each $g_\ell\,{\in}\,G$. 
This pairing is one-to-one because group multiplication is a bijection. 
Crucially, we assign the same hopping amplitude $t_{n,\ell}$ to all the hopping terms between sites $n$ and $\ell$. 
Different pillar-to-pillar bonds may carry different connections and amplitudes, and this is what allows non-trivial flux patterns. 

The single-particle Hilbert space is $\mathscr{H} = \bigoplus_{n} \mathscr{H}_n$ with $\mathscr{H}_n = \mathrm{span}\{|n, g\rangle : g \in G\}$, where $\ket{n,g}$ is the orbital labeled by group element $g$ at site $n$. Each pillar contributes $|G|$ basis states, and $N$ pillars generate a Hilbert space of $\dim \mathscr{H} = N \times |G|$. 
The hopping between pillars is then described by the CSL Hamiltonian
\begin{equation}
\label{eqn:CSL-Ham}
H = \sum_{n,\ell \,\in\, \textrm{sites}} \! t_{n\ell} \! \sum_{g,h \, \in  \, G} \! \;c^\dagger_{n,g} \, [\rho(a_{n,\ell})]_{g,h} \, c_{\ell,h}.  
\end{equation}
Here, $\rho(a_{n,\ell})$ is the regular representation~\cite{Serre:1977} of $G$, which describes the  $\abs{G}\times\abs{G}$ permutation matrix describing left multiplication by $a_{n,\ell}$. 
Specifically, we have $[\rho(a)]_{g,h} = 1$ if $g = a \cdot h$ and zero otherwise, so the sum over $g$ and $h$ picks out exactly the hoppings from $\ket{\ell, h}$ to $\ket{n, a_{n,\ell} \cdot h}$. 
For instance, $[\rho(+\mathrm{i})]_{+\mathrm{k},+\mathrm{j}} = 1$ because $+\mathrm{k} = (+\mathrm{i}) \cdot (+\mathrm{j})$. 
[See Supplementary Note~1 of the Supplementary Information for explicit constructions of the regular representations of $Z_2$ and $Q_8$.]
The resulting internal structure of pillars of group-labeled orbitals connected by regular representations is simple enough to enable universal feasibility in meta-material implementations. 

We point out that CSLs as considered in this work extend the earlier constructions of the same name~\cite{Lux2024,Marciani_2025} in two important aspects. 
First, we do not require the entirety of the orbitals (that is, all physical sites times the pillar content) to transform as elements of a single group, but merely that the construction respects a specified $\mathsf{SG}$ in a sense that will be specified shortly. 
Second---and crucially for the realization of topological models---we allow the hopping amplitudes $t_{n,\ell}$ to depend on the sites $\ell$ and $n$ in a manner that preserves~$\mathsf{SG}$.

\subsectitle{Synthetic gauge structure}
CSLs naturally give rise to synthetic gauge structures with the chosen group $G$, henceforth called the gauge group. 
Specifically, we can permute the orbitals within a pillar as $g_{n} \mapsto \chi_{n} \cdot g_{n} = \tilde{g}_{n}$ where $\chi_{n} \in G$ is the gauge transformation at site $n$ (see also Supplementary Note~3).
In the elementary case of gauge group $Z_2$, the non-trivial gauge transformation $\chi_{n} = -1$ exchanges the two orbitals at site ${n}$, which results in a change of the connection on bonds adjacent to ${n}$ (see lower two rows in Fig.~\ref{fig:CSL+GT}\bfsf{d}).
For a general choice of $G$ and $\chi$, the connection transforms as 
\begin{equation}
\label{eqn:connection-transformation}
\tilde{a}_{n,\ell} = \chi_{n} \cdot a_{n,\ell}\cdot \chi_{\ell}^{-1}.
\end{equation}

This gauge freedom allows us to reorder orbitals within a pillar and to apparently reshuffle the couplings between orbitals of adjacent pillars while still describing the same physical system. 
Nevertheless, certain characteristics remain invariant in this process. 
Specifically, the `integral' of connection on a closed loop $\gamma$ (assumed to be based at site ${n}$ and passing through sites $\ell_1,\ldots,\ell_{M}$), defined as
\begin{equation}
\label{eqn:Wilson-loop-def}
W_\gamma = a_{n,\ell_M}\cdot a_{\ell_M,\ell_{M-1}}\cdot \ldots \cdot a_{\ell_2,\ell_1}\cdot a_{\ell_1,n}, 
\end{equation}
transforms under a general choice of gauge transformation to $\tilde{W}_\gamma = \chi_{n} \cdot W_\gamma \cdot \chi_{n}^{-1}$.
It follows that the integral, called Wilson loop or flux on $\gamma$, is well-defined as a conjugacy class in the gauge group $G$. 
(Note that we here slightly deviate from the standard lattice gauge theory nomenclature~\cite{Goldman_2014} where these terms are instead reserved for the trace of $W_\gamma$.)
For Abelian groups like $Z_2$, each element of $G$ constitutes its own conjugacy class. 
For $Q_8$, one finds five conjugacy classes, namely $\{1\}$, $\{\pm\imi\}$, $\{\pm\imj\}$, $\{\pm\imk\}$ and $\{-1\}$, which were previously related to non-trivial braiding of vortex defects in biaxial nematic liquids~\cite{Poenaru:1977, Volovik:1977, Mermin:1979} and of band nodes in metals~\cite{Wu:2018, Bouhon2020, Bouhon:2020,Tiwari:2020, Bouhon2021mag,slager2024floquet,Jiang:2021}.
This noncommutativity also plays a pivotal role when introducing concrete models with quaternion fluxes in the sections below.

\subsectitle{Symmetry representations}
The inherent presence of a synthetic gauge structure implies that representations of symmetries in CSLs can be modified by gauge transformations. 
With this, we mean that simply translating the pillar locations with symmetry $s \in \mathsf{SG}$ from position vectors $\bs{r}_{n}$ to $s(\bs{r}_{n})$ may not preserve the arrangement of the connections; however, the connections could be restored if composing the geometric transformation with a gauge transformation $\chi$.
To illustrate the concept of gauge-modified symmetry, consider the one-dimensional chain with $G=Z_2$ and alternating connections $a_{n+1,n}=\pm 1$ shown in Fig.~\ref{fig:CSL+GT}\bfsf{d} (top row). 
Translation by one pillar (middle row) results in an exchange of the horizontal vs.~swapped bonds; however, applying a gauge transformation $\chi_{n} = -1$ at every second site (red rectangles) results in the restoration of the original connection (bottom row).
The presence of gauge-modified symmetries typically implies that $\mathsf{SG}$ is represented projectively.
Projectively represented $\mathsf{SG}\textrm{s}$ in lattices with synthetic $Z_2$ gauge have been explored over the past five years in one~\cite{Zhao:2021,Zhao_2024}, two~\cite{Zhao_2020,Zhao:2021,Li:2022,Chen:2022,Chen2023,Li2023acoustic,Lai2024phononic,Pu2023klein}, and three~\cite{Zhao:2021,ZHU2024,Chen:2025,teo2025observationembeddedtopologytrivial} spatial dimensions, resulting in a non-trivial extension of band theory.

We find that going beyond $G=Z_2$ introduces the additional option of automorphism-modified symmetries. 
In this context, automorphism $\varphi\in\mathrm{Aut}(G)$ of $G$ is an isomorphism of the group such that $\varphi(g_1\cdot g_2) = \varphi(g_1)\cdot \varphi(g_2)$.
A non-trivial example for $Q_8$ is the cyclic permutation of the imaginary units, 
\begin{equation}
\label{eqn:cyclic-imags}
\varphi_\textrm{cyclic}: +\imi \mapsto +\imj \mapsto +\imk \mapsto +\imi.
\end{equation} 
[The complete characterization of $\mathrm{Aut}(Q_8)$ and a graphical comparison of gauge transformations to gauge automorphisms appear in Supplementary Note~2 and 3.]
The key observation is that a global permutation of orbitals within all pillars with automorphism $\varphi$ ensures that the transformed CSL maintains a synthetic gauge structure with connection valued in $G$.
Therefore, one can attempt to compose the real-space translations $s\in \mathsf{SG}$ of the pillars with such a global permutation within the pillars to restore the original connection.
The possibility of automorphism-modified space group symmetries constitutes an additional avenue in band theory, and it naturally arises in the two models introduced in the next~pages.

\subsectitle{Peter-Weyl decomposition}
Owing to the interpretation of the orbitals within each pillar as elements of the gauge group, the dynamics of a CSL decouples into several blocks. 
To see this, let $\psi_{n,g}$ be the amplitude of a wave function at site $n$ in orbital $g$.
Gauge transformation by $\chi_{n}$ at site $n$ acts on the amplitudes by the regular representation~\cite{Serre:1977} $\rho(\chi_{n})$. 
A standard result known as the Peter-Weyl theorem~\cite{Sun:2024,Marciani_2025} ensures that the regular representation is reducible, with every irreducible representation (`irrep') $K$ of the group appearing a number of times equal to its dimension $d_K$.
Specifically, the decomposition 
\begin{equation}
\label{eqn:Peter-Weyl-decomp}
U^\dagger \rho(\chi_{n}) U = \bigoplus_K [D^K(\chi_{n})]^{\oplus d_K}    
\end{equation} 
is achieved with the unitary $U$ in Eq.~(\ref{eqn:unitary-Peter-Weyl}) of the Methods section, where $D^K(\chi)$ is the matrix representation of $\chi$ in irrep $K$.
We show in the Methods section that the connection $a_{n,\ell}$, by virtue of its compatibility with the group structure, couples the same irrep $D^K$ at various pillars. 
It therefore follows that the unitary $U$ results in block decomposition of the Hamiltonian of a CSL as well as of the Wilson operators~in~Eq.~(\ref{eqn:Wilson-loop-def}).

We investigate the implications of Eq.~(\ref{eqn:Peter-Weyl-decomp}) for gauge group $Q_8$. 
The quaternion group has five irreps: the trivial one-dimensional irrep ($A_0$; maps all elements to $+1$), three non-trivial one-dimensional irreps ($A_{x,y,z}$; each maps half of the elements to $+1$ and the other half to $-1 = e^{i\pi}$), and one two-dimensional irrep ($E$).
Higher-dimensional irreps are unique only up to unitary equivalence; here we specify the matrices $D^E(g)$ as
\begin{equation}
\label{eqn:E-irrep-matrices}
\begin{split}
D^E(\pm 1) = \pm\mathbb{1}, \,\;&\;\;\;\;
  D^E(\pm \imi) = \mp i \sigma_x, \\  
D^E(\pm \imj) = \mp i \sigma_y, &\,\;\;\;\;\;
  D^E(\pm \imk) = \mp i \sigma_z \, ,
\end{split}
\end{equation}
where $\sigma_{x,y,z}$ are the Pauli matrices, and $\mathbb{1}$ is the $2\times 2$ identity matrix.
The Hamiltonian therefore decouples into six sectors as $ U^\dagger H U = \bigoplus_K [H^K]^{\oplus d_K}$ with physical interpretation as follows. 
Each of the sectors $H^{A_{0,x,y,z}}$ supports a single (spinless) degree of freedom per pillar. 
Due to the reduction of the Wilson operators, sectors $H^{A_{x,y,z}}$ perceive an effective $U(1)$ flux quantized to $\phi\in\{0,\pi\}$ on every closed loop (thus replicating lattice structures with synthetic $Z_2$ gauge~\cite{Zhao_2020,Chen:2022,Chen2023,Zhao:2021,Zhao_2024,Chen:2025,Xue:2022,Li:2022,Li2023acoustic,Pu2023klein,Lai2024phononic,ZHU2024,teo2025observationembeddedtopologytrivial}), while fluxes are absent in sector $H^{A_0}$.

\begin{figure*}[t]
  \centering
  \includegraphics[width=0.99\linewidth]{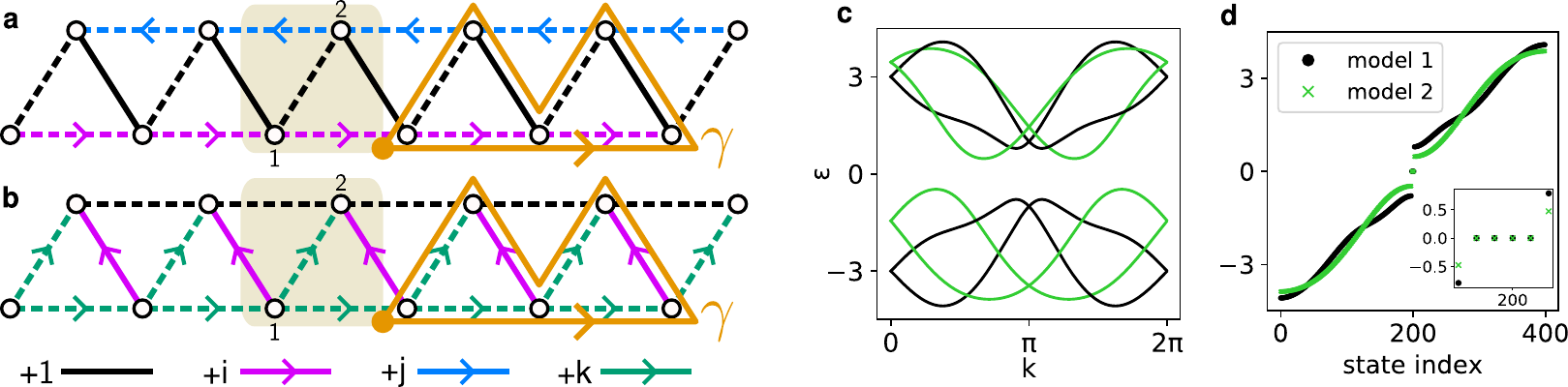}
  \caption{
  \textbf{Triangular ladders with quaternion 
  fluxes.}
  \bfsf{a,b} Translationally symmetric triangular ladders, both exhibiting flux $W_\triangle=\{\pm\imi\}$ on the lower and $W_\triangledown = \{\pm\imj\}$ on the upper set of triangles, with unit cells higlighted in beige. 
  Hopping amplitude is $t_1$ on the dashed ($t_2$ on the solid) bonds. 
  The coloring of the bonds indicates the connection $a_{n,\ell}$ according to the legend below panel~$\bfsf{b}$.
  The two models are distinguished by the flux $W_\gamma$ on path $\gamma$ (orange). 
  \bfsf{c,d} Energy spectra $\varepsilon(k)$ of the model in~$\bfsf{a}$ (black) and in~\bfsf{b} (green) with parameters $t_1 = 1$ and $t_2 = 2$ for the $H^E$ sector, in which the non-Abelian quaternion fluxes are faithfully represented.
  Both models have a gapped band structure in the bulk (panel \bfsf{c}) while finite chains support a pair of topological zero-energy modes on each boundary (panel \bfsf{d}).
  }
  \label{fig:nA-SSH}
\end{figure*}

In addition, the two-dimensional irrep $E$ results in two identical sectors, $H^{E,1} = H^{E,2} \equiv H^E$, each supporting a two-level degree of freedom per pillar. 
We reveal by virtue of concrete models that this sector models spin-$\frac{1}{2}$ electrons with spinful time-reversal symmetry, and elaborate further in the Methods section.
The connection $a_{n,\ell}$ in these sectors takes the values listed in Eq.~(\ref{eqn:E-irrep-matrices}), reminiscent of Rashba-type antisymmetric spin-orbit coupling for spinful electrons moving in a locally inversion-breaking environment~\cite{Kane:2005,Winkler:2003,Bauer:2012}. 
In this work, we adopt an alternative interpretation as spin-$\tfrac{1}{2}$ particle moving in $SU(2)$ gauge field whose flux on any closed loop is quantized to the right-hand sides of Eq.~(\ref{eqn:E-irrep-matrices}).

We thus find that a CSL with gauge group $Q_8$ naturally encompasses models of spinless particles moving on a lattice with quantized flux $Z_2 < U(1)$ (where `$<$' denotes a subgroup) as well as spinful particles with quantized flux $Q_8 < SU(2)$.
By default, the energetics of these sectors overlap. 
We present in Supplementary Note~4 a possible approach to splitting the energetics of the sectors by decorating each site with an on-site `pillar Hamiltonian'. 
As an experimentally practical alternative, we also discuss within the context of envisioned experimental simulations how a desired sector could be extracted even in the absence of the energy splitting through means of signal interference.
In the remainder of our work we therefore focus solely on the sector $H^{E}$ as it accommodates a faithful representation of the quaternion group, realizing a synthetic gauge structure with a non-Abelian gauge group.

\subsectitle{Triangular ladders with quaternion fluxes}

\noindent
To illustrate the concept of CSL with a non-Abelian synthetic gauge structure, we introduce two one-dimensional models with translation symmetry, both based on decorating a triangular ladder as shown in Fig.~\ref{fig:nA-SSH}\bfsf{a} and \bfsf{b}.
The unit cell of each model (beige region) contains two sites (white dots), each representing an eight-component pillar as in Fig.~\ref{fig:CSL+GT}\bfsf{c}.
The dashed (solid) bonds indicate hoppings of strength $t_1$ ($t_2$), and the connection is indicated with color (legend below Fig.~\ref{fig:nA-SSH}\bfsf{b}).
The Bloch Hamiltonians of the two models are given by Eqs.~(\ref{eqn:ladder-a-Ham}) and~(\ref{eqn:ladder-b-Ham}) of the Methods section.

Both models realize the same flux arrangement, with consecutive triangles characterized by alternating Wilson loops $W_\triangle = \{\pm \imi\}$ and $W_\triangledown = \{\pm \imj\}$.
However, the models are not related by a gauge transformation, which is revealed by comparing the Wilson loop along closed path $\gamma$ (orange) that encircles two corner-sharing triangles with the same flux $W_\triangle$.
In Fig.~\ref{fig:nA-SSH}\bfsf{a}, the Wilson loop along $\gamma$ is $W_\gamma = -1$, while $W_\gamma = +1$ in the case of Fig.~\ref{fig:nA-SSH}\bfsf{b}.
This distinction is clearly manifested by the different bulk energy spectra of the $\mathcal{H}^E(k)$ sector, with $\mathcal{H}$ the Bloch Hamiltonian and $k$ the one-dimensional momentum, plotted in Fig.~\ref{fig:nA-SSH}\bfsf{c}. 
That two fluxes $\{\pm \imi\}$ can combine into a net flux $+1$ or $-1$ is enabled by non-commutativity of the quaternion group~\cite{Poenaru:1977, Volovik:1977, Mermin:1979,Wu:2018, Bouhon2020, Bouhon:2020,Tiwari:2020, slager2024floquet,Jiang:2021}, and it illustrates that the connection (rather than the flux arrangement) is necessary to uniquely fix the gauge structure~\cite{Goldman_2014}.

We next use the triangular ladders to illustrate how the synthetic gauge forces $\mathsf{SG}$ symmetries to be modified by gauge and automorphisms.
We explicitly consider the 
model of Fig.~\ref{fig:nA-SSH}\bfsf{a} (discussion of the model in Fig.~\ref{fig:nA-SSH}\bfsf{b} is included in the Methods section and in Supplementary Note~6).
Adopting first $t_1 = t_2$, the model exhibits mirror symmetry 
\begin{equation}
\label{eqn:ladder-a-Mx}
    \mathsf{M}_x = \begin{bmatrix}  -i \sigma_z & \mathbb{0} \\ \mathbb{0} & -i \sigma_z \end{bmatrix} \begin{bmatrix}
        \mathbb{1} & \mathbb{0} \\ \mathbb{0} & e^{ik} \mathbb{1} 
    \end{bmatrix},
\end{equation}
that acts on the Bloch Hamiltonian as $\mathsf{M}_x(k) \mathcal{H}^E(k) \mathsf{M}_x^\dagger(k) = \mathcal{H}^E(-k)$.
The expression in Eq.~(\ref{eqn:ladder-a-Mx}) is interpreted as follows.
The right matrix corresponds to reflecting the site locations (with site $1$ kept in place and site $2$ moved by one unit cell to the left).
Looking at Fig.~\ref{fig:nA-SSH}\bfsf{a}, we find that the reflection reverts the sign of the connections $\pm\imi$ and $\pm\imj$ on the horizontal bonds.
This sign flip is corrected by virtue of Eq.~(\ref{eqn:connection-transformation}) with gauge transformation $\chi_1 = \chi_2 = +\imk$, implemented with the left matrix in Eq.~(\ref{eqn:ladder-a-Mx}).
The utilization of a gauge transformation in Eq.~(\ref{eqn:ladder-a-Mx}) means that $\mathsf{M}_x$ is gauge-modified. 

On the other hand, the model exhibits a glide reflection, which acts as $\mathsf{G}_y(k)\mathcal{H}^E(k)\mathsf{G}_y^\dagger(k) = \mathcal{H}^E(k)$, represented by
\begin{equation}
\label{eqn:ladder-a-Gy}
    \mathsf{G}_y= \begin{bmatrix} (\sigma_y - \sigma_x)/\sqrt{2} & \mathbb{0} \\ \mathbb{0} & (\sigma_y - \sigma_x)/\sqrt{2} \end{bmatrix} \begin{bmatrix} \mathbb{0} &  e^{i k} \mathbb{1} \\ \mathbb{1} & \mathbb{0} \end{bmatrix}.
\end{equation}
The right matrix in Eq.~(\ref{eqn:ladder-a-Gy}) again encodes the real-space action, with site 2 shifted to the location of site 1 in the next unit cell.
However, in contrast to Eq.~(\ref{eqn:ladder-a-Mx}), the left matrix in Eq.~(\ref{eqn:ladder-a-Gy}) does not correspond to a gauge transformation with elements in $Q_8$. 
Specifically, observe that the glide reflection replaces bonds with connections $\imi \leftrightarrow -\imj$, which cannot be corrected with a gauge transformation; however, the original connection is restored with the automorphism that maps $(+\imi,+\imj,+\imk)\mapsto(-\imj,-\imi,-\imk)$.
Adopting the representation of quaternions in Eq.~(\ref{eqn:E-irrep-matrices}), such automorphism is achieved through conjugation with the term `$(\sigma_y - \sigma_x)/\sqrt{2}$' that appears on the diagonal of the left matrix in Eq.~(\ref{eqn:ladder-a-Mx}).
Therefore, the glide reflection provides an example of an automorphism-modified symmetry.

We next generalize to $t_2 \neq t_1$. 
This breaks the $\mathsf{M}_x$ and $\mathsf{G}_y$ symmetries, but retains time-reversal, chiral, and 
$\pi$-rotation symmetries represented as $\mathcal{T} = \sigma_0 \otimes i \sigma_y \mathcal{K}$, $\mathcal{S} = \sigma_z \otimes \sigma_z$, and $\mathsf{C}_{2z} = \sigma_x \otimes i (\sigma   _x + \sigma_y) / \sqrt{2}$, respectively, where we used $\mathcal{K}$ for complex conjugation.
It is noteworthy that the time-reversal operator acquires the familiar spinful form, despite the spinlessness of the orbitals constituting the CSL. 
We discuss the origin of $\mathcal{T}^2 = - 1$ in the Methods section, with further clarifications included in Supplementary Note~5. 
Observe further that the $\pi$-rotation $\mathsf{C}_{2z}$ is automorphism-modified, with the second factor in the tensor product encoding the automorphism $(+\imi,+\imj,+\imk)\mapsto (+\imj,+\imi,-\imk)$.

We finally reveal that the triangular ladder with quaternion fluxes is topological.
Namely, the presence of spinful time-reversal ($\mathcal{T}^2 = -1$) that anticommutes with spinful rotation ($\mathsf{C}_{2z}^2= -1$) enables $Z_2$-valued partial polarization of Kramers-degenerate bands~(see Refs.~\onlinecite{Kane_2006}, \onlinecite{Ortix_2016}, and Supplementary Note~7).
We find that the partial polarization at half-filling is non-trivial for $\abs{t_2} > \abs{t_1}$, implying the appearance of a topological Kramers doublet inside the bulk energy gap at each edge, visible in the open-boundary spectra in Fig.~\ref{fig:nA-SSH}\bfsf{d}. 
The Kramers doublets are pinned to $\varepsilon=0$ by the chiral symmetry.

\begin{figure}[t]
  \centering
  \includegraphics[width=0.99\linewidth]{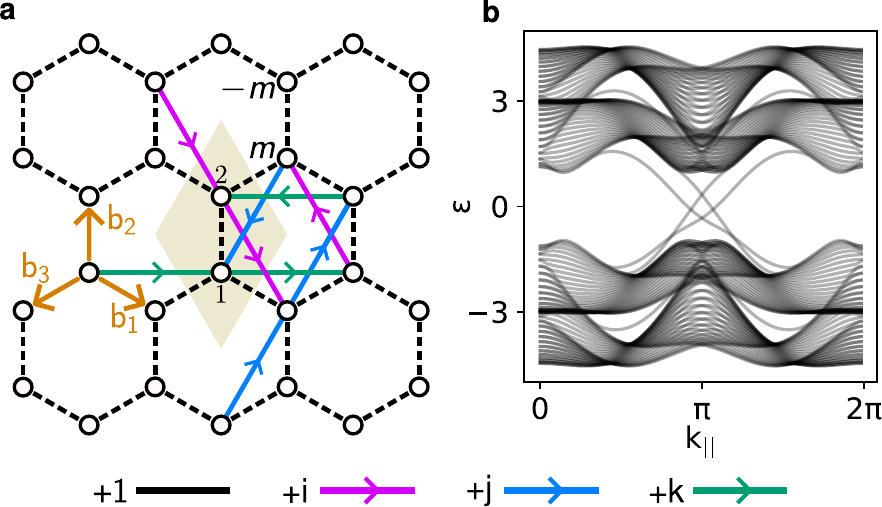}
  \caption{
  \textbf{Honeycomb model with quaternion fluxes.}
  \bfsf{a}~Model on the honeycomb lattice with nearest-neighbor (amplitude $t_1$, dashed lines) and next-nearest-neighbor hoppings (amplitude $t_2$, solid lines; only a few are explicitly shown). 
  Unit cell is displayed in beige.
  Orange arrows $\bs{b}_{1,2,3}$ indicate nearest-neighbor displacement vectors from sublattice `$1$' to sublattice `$2$'.
  The model exhibits quaternion-valued connection according to the color legend.
  \bfsf{b}~Spectrum $\varepsilon(k_\parallel)$ of a zigzag-terminated ribbon for parameters $t_1 = t_2 = m =1$ with helical edge modes inside the bulk energy~gap.
  } 
  \label{fig:nA-KM}
\end{figure}

\subsectitle{Honeycomb model with quaternion fluxes}

CSLs extend naturally to higher spatial dimensions. 
To demonstrate this, we construct a model on the honeycomb lattice with a non-Abelian synthetic gauge field as shown in Fig.~\ref{fig:nA-KM}.
The notation and the labeling follow that of Fig.~\ref{fig:nA-SSH}, and we additionally include the sublattice mass term $\pm m$. 
In the $H^E$ sector, with Bloch Hamiltonian given in Eq.~(\ref{eqn:honeycomb-Ham}), the non-Abelian gauge connection generates terms reminiscent of spin-orbit coupling, resulting in a Hamiltonian akin to the Kane-Mele model.~\cite{Kane:2005}
We probe the model's band topology by evaluating the Kane-Mele invariant using the Wilson-loop technique [see Code availability].~\cite{Soluyanov:2011,Gresch:2017} 
Setting $t_1 = 1$, we find that the band topology is non-trivial when $|t_2|>\sqrt{6} |m|$. 
In addition, we verify the existence of helical edge states in the topological phase, shown in Fig.~\ref{fig:nA-KM}\bfsf{b}. 
Owing to the presence of non-Abelian gauge structure, the model exhibits automorphism-modified symmetries.
For example, the sixfold symmetry $\mathsf{C}_{6z}$ around the center of a hexagon (whose representation is included in the Methods section) involves the cyclic automorphism in~Eq.~(\ref{eqn:cyclic-imags}).

\begin{figure}[t]
  \centering
  \includegraphics[width=0.9\linewidth]{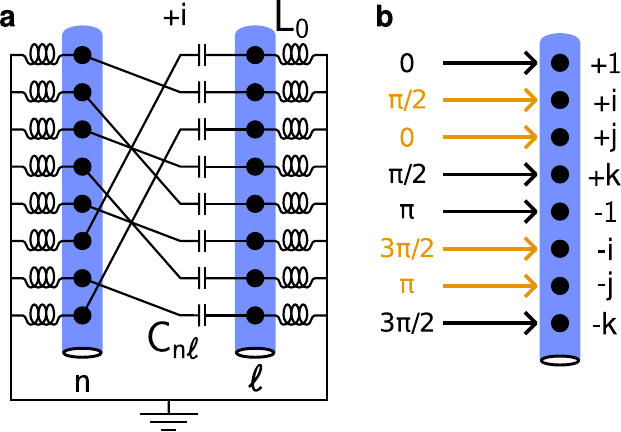}
  \caption{
  \textbf{Circuit implementation of a Cayley-Schreier lattice with quaternion fluxes.}
  \bfsf{a} 
  Every orbital of each pillar is replaced by a circuit node, and every hopping amplitude $t_{n,\ell}$ between pillars $n$ and $\ell$ is replaced with a capacitor $C_{n,\ell}$ while maintaining the connectivity dictated by the connection.
  Compare also the connection $a_{\ell,n}=+\imi$ here to $a_{n+1,n}=+\imi$ in Fig.~\ref{fig:CSL+GT}\bfsf{b}.
  In addition, all nodes are coupled to ground with an inductor $L_{0}$.
  \bfsf{b} 
  States transforming in the spinful irrep $E$ can be selectively excited with the knowledge of the Peter-Weyl decomposition in Eq.~(\ref{eqn:Peter-Weyl-decomp}) [also Eq.~(\ref{eqn:rotate-to-irrep-basis}) in Methods].
  To excite the state  $\ket{\uparrow}$ (state $\ket{\downarrow}$), one needs to inject AC current to the nodes marked with black arrows (with orange arrow) with relative phase shifts indicated to the~left.
  }
  \label{fig:circuit}
\end{figure}

\subsectitle{Blueprint for electric-circuit realization}

Experimental realization of CSLs with arbitrary gauge fields requires systems that can faithfully capture their inner structure. 
Electric-circuit networks---assemblies of resistors, inductors, and capacitors---provide an ideal platform, even enabling lattice geometries impossible in natural materials.~\cite{Lenggenhager:2021} 
The circuit Laplacian maps directly to tight-binding Hamiltonians~\cite{Lee:2018,Imhof:2018}, transforming standard electrical measurements into direct probes of band structures~\cite{Helbig:2019,Hofmann:2020}, density of states~\cite{Ningyuan:2015},
eigenstate profiles~\cite{Lenggenhager:2021}, and topological boundary modes~\cite{Imhof:2018}.
A particular advantage lies in controlling the boundary condition: direct edge wiring implements periodic boundaries whereas grounding of the edge sites creates open boundaries, with the specific choice potentially controllable within a single platform through a switch.~\cite{Yatsugi:2022} 
Alternatively, variable impedance elements based on operational amplifiers~\cite{Kotwal:2021} or analog multipliers~\cite{Chen:2023b} enable continuous tuning of the boundary condition.

Realization of CSLs in electric circuits is straightforward and proceeds according to the schematic in Fig.~\ref{fig:circuit}\bfsf{a}. Each orbital within a pillar becomes a node of the network, with nodes in adjacent pillars $n$ and $\ell$ connected via capacitors $C_{n,\ell}$ (proportional to the hopping amplitudes $t_{n,\ell}$) and with connectivity that mimics the connection $a_{n,\ell}$. 
In addition, each node in pillar $n$ couples to ground through inductor $L_n$ that sets the on-site mass (determining the frequency at which the bands form); choosing uniform $L_n \equiv L_0$ corresponds to the absence of sublattice potentials.

While bands from the different irrep sectors generically overlap, electric circuits enable selective excitation and probing of individual sectors through engineered current injection patterns.
Specifically, the Peter-Weyl decomposition [Eq.~(\ref{eqn:Peter-Weyl-decomp}), with the unitary $U$ for the quaternion group listed in Eq.~(\ref{eqn:rotate-to-irrep-basis})] specifies how every irrep corresponds to a specific linear combination of the orbitals within the pillars, thus dictating the relative amplitudes and phases of the injection currents necessary to excite states transforming in that irrep.
For example, the trivial irrep $A_0$ (first column of the matrix $U$) receives equal contributions from all group elements, necessitating in-phase injection currents of the same amplitude at all nodes. 
Conversely, irrep $A_y$ (third column of $U$) requires same-amplitude currents but with a relative $\pi$ phase shift between odd and even layers. 
To excite the two states in irrep $E$---corresponding in the spin-$\tfrac{1}{2}$ language to the states $\ket{\uparrow}$ and $\ket{\downarrow}$---we read columns $5$ and $6$ (or equivalently columns $7$ and $8$) of the $U$ in Eq.~(\ref{eqn:rotate-to-irrep-basis}).
For each of the two states, only half of the orbitals in the pillar receive an injection current, with their relative phases shown in Fig.~\ref{fig:circuit}\bfsf{b} (orange for state $\ket{\uparrow}$, black for state $\ket{\downarrow}$).

With the relative phase shifts and amplitudes of the injection currents of the desired irrep determined, probing of the spectrum corresponds to electrical measurements at variable drive frequency $\omega$.
Specifically, the ground impedance $Z(\omega) = V(\omega)/I(\omega)$ exhibits sharp resonances at eigenmode frequencies, providing direct spectral mapping.
In the triangular ladder CSL realization (Fig.~\ref{fig:nA-SSH}\textbf{a}), topological edge modes manifest as isolated peaks within the bulk gap. 
The Kramers degeneracy of the edge mode implies that the impedance is the same for injection currents corresponding to both states $\ket{\uparrow}$ and $\ket{\downarrow}$.
Lastly, to measure the wave function of the edge modes, one simply tunes the drive to the mode’s resonance frequency and measures the spatial profile of the induced voltage.

%%%%%%%%%%%%%%%%%%%
%%% Discussion  %%%
%%%%%%%%%%%%%%%%%%%

\sectitle{Discussion}
We showed that Cayley-Schreier lattices allow one to construct crystalline structures with space group symmetry while naturally incorporating non-Abelian fluxes. 
In addition, these models can be decomposed into sectors that correspond to pseudospin-$N$ particles coupled to quantized $U(2N\,{+}\,1)$ fluxes, with $N$ set by the dimension of the sector. 
Some of these pseudospins correspond to true spinors; in particular, the quaternion group $Q_8$ is the smallest gauge group that enables the simulation of spin-$\frac{1}{2}$ fermions.
Each of these (pseudo-)spin models can, in principle, realize any kind of band topology within a single setup. 
We have demonstrated this by formulating representative models in one and two spatial dimensions that simultaneously exhibit both spinless and spinful excitations. 
Due to the simple internal structure of CSLs, the construction of these models in fact only requires elementary hopping terms, making them suitable for meta-material realizations as we have also detailed.

The above results set the stage for multifold follow-up directions, both theoretically and experimentally. 
In this regard, an immediate pursuit is to realize the presented models, as well as generalizations thereof, in electric-circuit setups or other meta-material platforms along the specific details outlined. 
From a theoretical angle, our work invites a broad range of generalizations. 
To appreciate the prospects of CSLs, note that the inclusion of $Z_2$ fluxes in crystalline structures---corresponding to the simplest choice of gauge theory---has led to a substantial body of work in topological band theory~\cite{Zhao_2020,Zhao:2021,Chen:2022,Chen2023,Zhao_2024,Chen:2025,Xue:2022,Li:2022,Li2023acoustic,Pu2023klein,Lai2024phononic,ZHU2024,teo2025observationembeddedtopologytrivial}, which includes the discovery of Möbius topological insulators~\cite{Li2023acoustic}, Klein Brillouin zones~\cite{Chen:2022,Pu2023klein}, shifting of high-symmetry points~\cite{Chen2023}, and the formation of platycosms in the momentum space~\cite{Chen:2025}, and whose topological implications continue to be investigated at present.
While all these phenomena can be reproduced by CSLs by setting the gauge group to $G\,{=}\,Z_2$, the presented platform has the notable flexibility of implementing arbitrary finite gauge groups $G$---including those with non-trivial automorphisms $\varphi\in\textrm{Aut}(G)$ and even non-Abelian ones---which we anticipate to vitally broaden the scope of topological band theory.
In particular, the presented models are directly generalizable to gauge structures supporting higher-spin particles, which lack simple equivalents in electronic and photonic band structures.
Besides topological insulators, we anticipate such high-spin particles to enable exotic types of band degeneracies and semimetals.
The presented models also call for a systematic characterization of non-Abelian gauge structures even if narrowing attention to the quaternion group: 
to enumerate symmetry-compatible synthetic gauge structures for all wallpaper groups and space groups on the one hand, and to classify topological bands in every such gauge arrangement~on~the~other. 

Finally, our work also raises additional structural questions. 
Indeed, the CSL construction invites refined mathematical methods to capture non-Abelian synthetic gauge structures and CSLs, because conventional methods such as group cohomology evaluations~\cite{Chen:2013,Alexandradinata:2016,Thorngren:2018} are specific to Abelian gauge groups. 
Additionally, such analyses do not include symmetries generated by automorphisms of the gauge group, suggesting that a more general framework is at play. 
Not in the least place, taking the non-Abelian CSL identified so far and considering quantum and correlated phases as well as disorder in this geometry, in which case all sectors interplay, sets yet another intriguing direction to consider.

%%%%%%%%%%%%%%%%
%%% METHODS  %%%
%%%%%%%%%%%%%%%%

\sectitle{Methods}

\subsectitle{Block-diagonalization of CSL Hamiltonians}
The Peter-Weyl decomposition of CSL Hamiltonians is achieved via a unitary rotation in the orbital basis. 
The original basis, where each basis vector corresponds to a group element $g \in G$, is transformed to a basis labeled by the irreducible representations of the gauge group.
Following Ref.~\citenum{Sun:2024}, the transformation is defined as 
\begin{equation}
\label{eq:annihilation-rotation}
\left[\tilde{\psi}^\dagger\right]_\mu =  \tilde{c}^\dagger_\mu = \sum_g U_{g,\mu}  c^\dagger_g = \sum_g U_{g,\mu} \left[\psi^\dagger\right]_g ,
\end{equation}
where the $\left[\tilde{\psi}\right]_\mu = \tilde{c}_\mu$ are components of the annihilation operator vector in the irrep basis and $U_{g,\mu}$
are the components of the unitary matrix of the basis transformation. 
We introduced $\mu$ as a composite index representing the different $(K,\lambda,\nu)$ tuples, where $K$ denotes an irreducible representation of $G$, $\lambda$ enumerates the occurrence of $K$ in the decomposition of the regular representation, and $\nu \in {1,\dots,d_K}$ labels the basis states within the irrep of dimension $d_K$. 
In this way, Eq.~(\ref{eq:annihilation-rotation}) is compactly written as $\tilde{\psi}^\dagger = \psi^\dagger U$.
Generally, the matrix $U$ is given by~\cite{Sun:2024} 
\begin{equation}
\label{eqn:unitary-Peter-Weyl}
U_{g,(K,\lambda,\nu)} = \sqrt{\frac{d_{K}}{\abs{G}}} \left[D_{\nu\lambda}^K(g)\right]^*,
\end{equation}
where $D^K_{\nu\lambda}(g)$ are the matrix elements of the irrep, and the star denotes complex conjugation.

CSL Hamiltonians are kept invariant under such unitary transformations, since each quadratic term is invariant,
\begin{equation}
    \psi^\dagger_{n} \rho(a_{n,\ell}) \psi_{\ell} = \psi_{n}^\dagger U U^\dagger \rho(a_{n,\ell}) U U^\dagger \psi_{\ell} = \tilde{\psi}^\dagger_{\ell} \tilde{\rho}(a_{n,\ell}) \tilde{\psi}_{\ell},
\end{equation}
where $\rho(a_{n,\ell})$ is the regular representation of the connection that encodes hopping amplitudes from site ${\ell}$ to site $n$. 
Importantly, rotation to the irrep basis $\tilde{\rho}(\chi_{n,\ell}) = U^\dagger \rho(\chi_{n,\ell}) U$ results in a block-diagonal form given by the decomposition in Eq.~(\ref{eqn:Peter-Weyl-decomp}).
Since the tight-binding Hamiltonian of CSLs generally takes the form given in Eq.~(\ref{eqn:CSL-Ham}),
that is, as a sum of regular representations of connections $a_{n,\ell}$ over all bonds, the entire CSL Hamiltonian is block-diagonalized by $1\otimes U$, where the identity acts on the site degree of freedom and $U$ acts on the orbitals within pillars.  

For the specific choice $G = Q_8$, and given the ordering of the group elements as $(+1, +\imi, +\imj,+\imk,-1,-\imi,-\imj,-\imk)$ in accordance with Fig.~\ref{fig:CSL+GT}\bfsf{b}, the matrix $U$ takes the form
\begin{equation}
\label{eqn:rotate-to-irrep-basis}
    U=
    \frac{1}{\sqrt{8}}
    \begin{bmatrix}
        1 & 1 & 1 & 1 & \sqrt{2} & 0 & 0 & \sqrt{2} \\
        1 & 1 & -1 & -1 & 0 & \sqrt{2} i  & \sqrt{2} i  & 0 \\
        1 & -1 & 1 & -1 & 0 & \sqrt{2} & -\sqrt{2} & 0 \\
        1 & -1 & -1 & 1 & \sqrt{2} i  & 0 & 0 & -\sqrt{2} i  \\
        1 & 1 & 1 & 1 & -\sqrt{2} & 0 & 0 & -\sqrt{2} \\
        1 & 1 & -1 & -1 & 0 & -\sqrt{2} i  & -\sqrt{2} i  & 0 \\
        1 & -1 & 1 & -1 & 0 & -\sqrt{2} & \sqrt{2} & 0 \\
        1 & -1 & -1 & 1 & -\sqrt{2} i  & 0 & 0 & \sqrt{2} i 
    \end{bmatrix}\!.
\end{equation}
In Eq.~(\ref{eqn:unitary-Peter-Weyl}), the subscripts $g$ and $(K,\lambda,\nu)$ correspond to the rows resp.~columns of the displayed unitary.

\subsectitle{Spinless vs.~spinful time-reversal symmetry}

We clarify how spinful time-reversal symmetry (in the sense that $\mathcal{T}^2 = -1$) emerges in the Hamiltonian block $\mathcal{H}^E$ of a CSL with gauge group $G = Q_8$. 
This observation may seem unexpected since the time-reversal operator $T$ at the level of the constituent orbitals (black dots in Fig.~\ref{fig:CSL+GT}) is manifestly spinless (meaning $T^2 = +1$). 
Specifically, it acts on a general wave function through complex conjugation,
\begin{equation}
T\Big[\sum_{n,g} \psi_{n,g} \ket{n,g}\Big] = \sum_{n,g} \psi_{n,g}^* \ket{n,g},
\end{equation}
where $n$ labels physical sites and $g$ the orbitals within pillars.

To understand how this time-reversal operator acts at the level of individual irreps, we study how it is changed under the unitary transformation in Eq.~(\ref{eqn:Peter-Weyl-decomp}). 
We find
\begin{equation}
\label{eqn:rotated-micro-TRS}
U^\dagger T U = U^\dagger U^* \mathcal{K} = 
\begin{bmatrix}
 1 & 0 & 0 & 0 & 0 & 0 & 0 & 0 \\
 0 & 1 & 0 & 0 & 0 & 0 & 0 & 0 \\
 0 & 0 & 1 & 0 & 0 & 0 & 0 & 0 \\
 0 & 0 & 0 & 1 & 0 & 0 & 0 & 0 \\
 0 & 0 & 0 & 0 & 0 & 0 & 0 & 1 \\
 0 & 0 & 0 & 0 & 0 & 0 & \!\!-1\! & 0 \\
 0 & 0 & 0 & 0 & 0 & \!\!-1\! & 0 & 0 \\
 0 & 0 & 0 & 0 & 1 & 0 & 0 & 0 \\
\end{bmatrix}\mathcal{K} , 
\end{equation}
which is not block-diagonal for the two-dimensional irrep: the off-diagonal form of the bottom-right $4\times 4$ elements implies that the two copies of irrep $E$ are~swapped under $T$.

However, the situation can be rectified. 
Since both copies of $E$ are associated with the same representations $D^E(g)$ and identical Hamiltonians $H^E$, we can compose $U^\dagger T U$ with a unitary matrix $M$ that mixes the two copies of $E$ while keeping the Hamiltonian blocks invariant. 
Specifically, choosing
\begin{equation}
M = \begin{bmatrix}
1 & 0 & 0 & 0 & 0 & 0 & 0 & 0 \\
0 & 1 & 0 & 0 & 0 & 0 & 0 & 0 \\
0 & 0 & 1 & 0 & 0 & 0 & 0 & 0 \\
0 & 0 & 0 & 1 & 0 & 0 & 0 & 0 \\
0 & 0 & 0 & 0 & 0 & 0 & \!\!-1\! & 0 \\
0 & 0 & 0 & 0 & 0 & 0 & 0 & \!\!-1\! \\
0 & 0 & 0 & 0 & 1 & 0 & 0 & 0 \\
0 & 0 & 0 & 0 & 0 & 1 & 0 & 0 \\
\end{bmatrix},
\end{equation}
we find that the construction of $\mathscr{T} = M \, U^\dagger T U$ results in
\begin{equation}
\label{emergent-big-TRS}
\mathscr{T} = 
\begin{bmatrix}
 +1\! & 0 & 0 & 0 & 0 & 0 & 0 & 0 \\
 0 & \!\!+1\! & 0 & 0 & 0 & 0 & 0 & 0 \\
 0 & 0 & \!\!+1\! & 0 & 0 & 0 & 0 & 0 \\
 0 & 0 & 0 & \!\!+1\! & 0 & 0 & 0 & 0 \\
 0 & 0 & 0 & 0 & 0 & \!\!+1\! & 0 & 0 \\
 0 & 0 & 0 & 0 & \!\!-1\! & 0 & 0 & 0 \\
 0 & 0 & 0 & 0 & 0 & 0 & 0 & \!\!+1\! \\
 0 & 0 & 0 & 0 & 0 & 0 & \!\!-1\! & 0
\end{bmatrix}
\mathcal{K} = \bigoplus_K [\mathcal{T}^K]^{\oplus d_K}.    
\end{equation}
We find the familiar spinful time-reversal operator $\mathcal{T}^E = i \sigma_y \mathcal{K}$ for the blocks transforming in the two-dimensional irrep.

For a general choice of gauge group $G$ and irrep $K$, the character of the emergent time-reversal symmetry $\mathcal{T}^K$ of the Hamiltonian $H^K$ is dictated by the Frobenius-Schur indicator~\cite{Serre:1977} 
\begin{equation}
\label{eqn:Frobenius-Schur}
\iota_K = \frac{1}{|G|}\sum_{g\in G} \tr[D^K(g^2)].
\end{equation}
This indicator always satisfies $\iota_K \in \{0,+1,-1\}$; specifically: 
\begin{enumerate}
    \item $\iota_K = 0$ implies that $H^K$ has no time-reversal symmetry,
    \item $\iota_K = +1$ implies spinless time-reversal symmetry, and 
    \item $\iota_K = -1$ implies spinful time-reversal symmetry.
\end{enumerate}
We derive this trichotomy of time-reversal operators in Supplementary Note~5.
Generally, odd-dimensional irrep $D^K$ cannot have a spinful $\mathcal{T}^K$; on the other hand, even-dimensional irrep can have either spinless or spinful $\mathcal{T}^K$.
We also remark that the quaternion group $Q_8$, as considered in this manuscript, constitutes the smallest gauge group that results in a CSL exhibiting spinful Hamiltonian blocks.

\subsectitle{Details of the triangular ladder model in Fig.~\ref{fig:nA-SSH}\bfsf{a}}

We write the Bloch Hamiltonian of both triangular ladders in the `periodic Bloch convention', which disregards the embedding of the sites within the unit cell.~\cite{Bena:2009}
The Bloch Hamiltonian of the model in Fig.~\ref{fig:nA-SSH}\bfsf{a} reads
\begin{equation}
\label{eqn:ladder-a-Ham}
    \mathcal{H}^E(k)= \begin{bmatrix}  
    2 t_1 \sin{k}\,\sigma_x  &
    (t_1+t_2e^{ik})\mathbb{1} \\
    (t_1+t_2e^{-ik})\mathbb{1} & 
    -2 t_1 \sin {k}\, \sigma_y \end{bmatrix}.
\end{equation}
Below, we supplement the discussion of the model's symmetries beyond the treatment given in the Results section.
In particular, we clarify why setting $t_1 = t_2$ enforces a gap closing at momentum $k=\pi$. 
This band touching, visible in Supplementary Fig.~1\bfsf{a}, signals a topological change of the partial polarization.

Observe that setting $t_1 = t_2$ leads to the appearance of the glide reflection $\mathsf{G}_y$ [Eq.~(\ref{eqn:ladder-a-Gy})], which squares to translation by one unit cell.
It follows that all bands carry eigenvalues $\lambda(k)=\pm e^{i k/2}$ of the glide symmetry.
For convenience, we color the energy bands in Supplementary Fig.~1\bfsf{a}, according to the complex phase of their glide eigenvalue.
Crucially, the fraction $1/2$ in the eigenvalue exponent implies~\cite{Zhao:2016} that energy bands appear in glide doublets, visible in Supplementary Fig.~1\bfsf{a}, that are connected together across the Brillouin zone boundary.
The glide eigenvalues of the two bands within a glide doublet differ by the $\pm$ sign; specifically, $\lambda(0) = \pm 1$ are real while $\lambda(\pi) = \pm \imi$ are imaginary. 

We finally utilize the presence of spinful time-reversal symmetry, which enforces Kramers pairs at time-reversal-invariant momenta (TRIMs) $k\,{\in}\,\{0,\pi\}$.
The glide reflection and the time-reversal operators obey $\{\mathsf{G}_y,\mathcal{T}\}\,{=}\,0$ at both TRIMs, which in combination with the antiunitarity of $\mathcal{T}$ implies that the glide eigenvalues of the two states within a Kramers pair are $(\lambda,-\lambda^*)$.
Observe that the eigenvalues $(+1,-1)$ of a glide doublet at $k=0$ are compatible with the eigenvalues of a Kramers pair.
In contrast, the glide eigenvalues $(+\imi,-\imi)$ of a glide doublet at $k=\pi$ are not compatible with a Kramers pair.  
It follows~\cite{Bzdusek:2016} that, for $t_1 = t_2$, the four bands are tied together through a degeneracy of the central two bands.
In the presented model, this gap closing occurs in the form of four-fold degeneracy at momentum $k=\pi$, pinned by the chiral symmetry $\mathcal{S}$ to energy $\varepsilon=0$.

\subsectitle{Details of the triangular ladder model in Fig.~\ref{fig:nA-SSH}\bfsf{b}}

We next investigate the model in Fig.~\ref{fig:nA-SSH}\bfsf{b}, whose Bloch Hamiltonian
\begin{equation}
\label{eqn:ladder-b-Ham}
    \mathcal{H}^E(k)= \begin{bmatrix}  
    2 t_1 \sin{k}\,\sigma_z  &
    i (t_1\sigma_z +t_2e^{ik}\sigma_x)\\
    -i (t_1\sigma_z +t_2e^{-ik}\sigma_x) & 
    2 t_1 \cos {k}\, \mathbb{1} \end{bmatrix}
\end{equation}
produces the green bands in Fig.~\ref{fig:nA-SSH}\bfsf{c}.
While flux through the triangles implies that this nearest-neighbor model has chiral symmetry, representation of this symmetry requires unit-cell doubling.
We thus consider a four-site unit cell consisting of sites $1$ and $2$ in Fig.~\ref{fig:nA-SSH}\bfsf{b} together with their translations $1'$ and $2'$ to the right.
The folded Bloch Hamiltonian reads
\begin{equation}
\small
\label{eqn:ladder-b-Ham-doubled}
    \!\mathcal{H}^E_{2\times}\!(k)\!=\! t_1\!\!\begin{bmatrix} 
    \mathbb{0}  &
    i \sigma_z &
    \!i(1{-}e^{ik})\sigma_z\! &
    i \tau e^{ik} \sigma_x \\
    -i\sigma_z &
    \mathbb{0} &
    -i \tau \sigma_x &
    \!(1{+}e^{ik})\mathbb{1} \\
    i(e^{-ik}{-}1)\sigma_z \!&
    i \tau \sigma_x &
    \mathbb{0} &
    i \sigma_z \\
    \!-i \tau e^{-ik} \sigma_x\! &
    \!(1{+}e^{-ik})\mathbb{1}\! &
    -i \sigma_z &
    \mathbb{0}
    \end{bmatrix}\!
\end{equation}
where $\tau = t_2/t_1$.
Chiral symmetry is represented as $\mathcal{S}=\sigma_z\otimes \sigma_z \otimes \sigma_z$, and the model exhibits spinful time-reversal symmetry
$\mathcal{T} = \sigma_0\otimes \sigma_0 \otimes i\sigma_y \mathcal{K}$.
We show the band structure of the folded Bloch Hamiltonian for $t_1 = 1$ and $t_2 = 2$ in~Supplementary Fig.~1\bfsf{b}.

Note that the triangular ladder model in Fig.~\ref{fig:nA-SSH}\bfsf{b} is not symmetric under the operation of a glide plane $\mathsf{G}_y$ even if setting $t_1 = t_2$; 
this is because two consecutive fluxes $\{\pm \imi\}$ combine to $W_\gamma=+1$ while two consecutive fluxes $\{\pm \imj\}$ combine to $W_{\mathsf{G}_y\gamma}=-1$.
For the same reason, the $\mathsf{C}_{2z}$ symmetry is also absent. 
By contrast, setting $t_1 = t_2$ enables a gauge-modified mirror symmetry
\begin{equation}
\label{eqn:ladder-b-Mx}
    \mathsf{M}_x = \begin{bmatrix}  -i \sigma_x & \mathbb{0} & \mathbb{0} & \mathbb{0} \\ 
    \mathbb{0} & -i \sigma_z & \mathbb{0} & \mathbb{0} \\
    \mathbb{0} & \mathbb{0} & -i \sigma_x & \mathbb{0} \\
    \mathbb{0} & \mathbb{0} & \mathbb{0} &  -i \sigma_z
    \end{bmatrix} \begin{bmatrix}
    \mathbb{0} & \mathbb{0} & \mathbb{1} & \mathbb{0} \\ 
    \mathbb{0} & \mathbb{1} &\mathbb{0} & \mathbb{0} \\
    \mathbb{1} & \mathbb{0} & \mathbb{0} & \mathbb{0} \\
    \mathbb{0} & \mathbb{0} & \mathbb{0} &  e^{ik}\mathbb{1}
    \end{bmatrix},
\end{equation}
which involves gauge transformation $\chi_1 = \chi_{1'} = +\imi$ and $\chi_2=\chi_{2'} = +\imk$ on the four sites within the doubled unit cell.

On the other hand, translation by a single (not doubled) unit cell has to be treated as an additional symmetry of the folded Bloch Hamiltonian.~\cite{Zhao_2020}
This translation, which is a symmetry of the model irrespective of the choice of $t_1$ and $t_2$, is represented as 
\begin{equation}
\label{eqn:ladder-b-Tx}
\mathsf{L}_x=\begin{bmatrix}  \mathbb{0} & \mathbb{0} & e^{ik} \mathbb{1} & \mathbb{0} \\ 
    \mathbb{0} & \mathbb{0} & \mathbb{0} & e^{ik} \mathbb{1} \\
    \mathbb{1} & \mathbb{0} & \mathbb{0} & \mathbb{0} \\
    \mathbb{0} & \mathbb{1} & \mathbb{0} & \mathbb{0}
    \end{bmatrix}.
\end{equation}
and it acts on the Hamiltonian as $\mathsf{L}_x(k) \mathcal{H}^E_{2\times}\!(k) \mathsf{L}^\dagger_x(k) = \mathcal{H}^E_{2\times}\!(k)$ while squaring to $e^{ik}\mathbb{1}$.
We observe by comparing against Eq.~(\ref{eqn:ladder-a-Gy}) that the translation symmetry of ladder $\bfsf{b}$ acts on the Hamiltonian~(\ref{eqn:ladder-b-Ham-doubled}) in an analogous fashion as the glide symmetry of ladder $\bfsf{a}$ acts on the Hamiltonian~(\ref{eqn:ladder-a-Ham}).
In particular, energy bands appear in translation doublets which are connected together across the Brillouin zone boundary and that carry translation eigenvalues $(e^{ik/2},-e^{ik/2})$.~\cite{Zhao:2016}

We include an extended analysis of the symmetry of the triangular ladder model of Supplementary Fig.~1\bfsf{b} in Supplementary Note~6.
Therein, we show how the combined effect of symmetries $\mathsf{L}_x$ and $\mathcal{T}$ for $t_1 \neq t_2$ enforces the butterfly-like spectrum shown in Supplementary Fig.~1\bfsf{b}, and how the combined effect of the set $\{\mathsf{L}_x$, $\mathcal{T}$, $\mathcal{M}_x, \mathcal{S}\}$ for $t_1 = t_2$ enforces a four-fold degeneracy at momentum $k=\pi$ as shown in Supplementary Fig.~1\bfsf{c}.

\subsectitle{Details of the honeycomb model}

We write the Bloch Hamiltonian of the quaternion-flux honeycomb model in the `non-periodic Bloch convention' which includes the embedding of the sites within the unit cell.~\cite{Bena:2009}
We set the nearest-neighbor distance to $1$, so that the three displacement vectors from sublattice `$1$' to sublattice `$2$' (shown in Fig.~\ref{fig:nA-KM}) are $\bs{b}_1 = (\sqrt{3},-1)/2$, $\bs{b}_2 = (0,1)$, and $\bs{b}_{3} =  (-\sqrt{3},-1)/2$, and we introduce their differences (Bravais vectors) $\bs{a}_1 = \bs{b}_1 - \bs{b}_2$, $\bs{a}_2 = \bs{b}_2 - \bs{b}_3$, and $\bs{a}_3 = \bs{b}_3 - \bs{b}_1$.
The Hamiltonian can then be expressed as 
\begin{eqnarray}
\label{eqn:honeycomb-Ham}
\mathcal{H}^E(\bs{k}) 
\!\!&=&\!\! \sigma_z \,{\otimes} \left[ m \,  \sigma_0 - 2t_2 \!\sum_{l = 1}^3 \sin(\bs{k} {\cdot} \bs{a}_l)\, \sigma_\ell \right]  \\
\!\!&\phantom{=}&\!\! + \,t_1 \! \left[\sigma_1 \! \sum_{l = 1}^3 \cos(\bs{k}{\cdot} \bs{b}_l) + \sigma_2 \! \sum_{l = 1}^3 \sin(\bs{k}{\cdot} \nonumber\bs{b}_l)\right] {\otimes} \,\sigma_0 
\end{eqnarray}
where $t_1$ ($t_2$) is the nearest-neighbor (next-nearest-neighbor) hopping amplitude and $m$ is the sublattice mass.

The constructed model with non-Abelian fluxes is invariant under the space group of the honeycomb lattice with gauge-modified and automorphism-modified symmetries.
For example, rotation by $2\pi/6$ around the center of a hexagon (present when the sublattice mass is set to $m=0$) permutes the connection on next-nearest-neighbor bonds as $+\imi \mapsto +\imk \mapsto +\imj \mapsto \imi$. 
Correcting for this change of the connection demands that the rotation is composed with the inverse automorphism $+\imi \mapsto +\imj \mapsto +\imk \mapsto + \imi$, represented with the matrix $V = (\mathbb{1} - i \sum_{l = 1}^3 \sigma_l)/2$ in the complement $SU(2)\backslash Q_8$. 
Therefore, the rotation is automorphism-modified as $\mathsf{C}_{6z} = \sigma_x \otimes V$ with the first factor coming from the exchange of the two sublattices under the symmetry and the second factor encoding the non-trivial automorphism.

%%%%%%%%%%%%%%%%%%%%%%
%%% DATA and CODE  %%%
%%%%%%%%%%%%%%%%%%%%%%

\sectitle{Data availability}
All the data used to arrive at the conclusions presented in this work are publicly available in the following data repository: \href{https://doi.org/10.5281/zenodo.18954513}{https://doi.org/10.5281/zenodo.18954513}.

\sectitle{Code availability}
All the Python code used to generate and analyze the data and to arrive at the conclusions presented in this work is publicly available in the form of annotated notebooks in the following data repository: \href{https://doi.org/10.5281/zenodo.18954513}{https://doi.org/10.5281/zenodo.18954513}.

%%%%%%%%%%%%%%%%%%%
%%% REFERENCES  %%%
%%%%%%%%%%%%%%%%%%%

\end{bibunit}
\let\oldaddcontentsline\addcontentsline     
\renewcommand{\addcontentsline}[3]{}        
\bibliography{ref}
\let\addcontentsline\oldaddcontentsline

%%%%%%%%%%%%%%%%%%%%%%%%
%%% ACKNOWLEDGMENTS  %%%
%%%%%%%%%%%%%%%%%%%%%%%%

\sectitle{Acknowledgments}
We would like to thank A.~Iliasov, M.~Marciani, T.~Neupert, and A.~Tiwari for valuable discussions.
Z.G., L.K.U, and T.B.~were supported by the Starting Grant No.~211310 by the Swiss National Science Foundation. R.-J.S.~acknowledges funding from an EPSRC ERC underwrite grant  EP/X025829/1, and a Royal Society exchange grant IES/R1/221060.

%%%%%%%%%%%%%%%%%%%%%%
%%% CONTRIBUTIONS  %%%
%%%%%%%%%%%%%%%%%%%%%%

\sectitle{Author contributions}
L.K.U.~and T.B.~conceived the project.
Z.G.~and T.B.~analyzed the presented tight-binding models.
R.-J.S.~advised on the gauge theory aspects.
L.K.U.~and T.B.~devised the electric-circuit blueprint.
All authors contributed to developing the theoretical framework, discussed together, and wrote the manuscript.

%%%%%%%%%%%%%%%%%%%%%%%%%%%%
%%% COMPETING INTERESTS  %%%
%%%%%%%%%%%%%%%%%%%%%%%%%%%%

\sectitle{Competing interests}
The authors declare no competing interests.

\end{document}

\begin{bibunit}
\onecolumngrid
\renewcommand\thesection{\Roman{section}}
\renewcommand\thesubsection{\Alph{subsection}}
\setcounter{page}{1}

\setcounter{figure}{0}
\setcounter{equation}{0}
\setcounter{table}{0}

\makeatletter
\c@affil=0
\makeatother

\renewcommand{\theequation}{S\arabic{equation}}
\renewcommand*{\theHsection}{\the\value{section}}

\title{Supplementary Information for: \texorpdfstring{\medskip \\}{} 
Topological non-Abelian gauge structures in Cayley-Schreier lattices}

\author{Zoltán Guba\,\orcidlink{0000-0002-6130-1064}}
\affiliation{Department of Physics, University of Z\"urich, Winterthurerstrasse 190, 8057 Z\"urich, Switzerland}
\author{Robert-Jan Slager\,\orcidlink{0000-0001-9055-5218}}
\affiliation{Department of Physics and Astronomy, University of Manchester,
Oxford Road, Manchester M13 9PL, United Kingdom}
\author{Lavi K.~Upreti\,\orcidlink{0000-0002-1722-484X}}\email{lavi.upreti@uzh.ch}
\affiliation{Department of Physics, University of Z\"urich, Winterthurerstrasse 190, 8057 Z\"urich, Switzerland}
\author{Tom\'a\v{s} Bzdu\v{s}ek\,\orcidlink{0000-0001-6904-5264}}\email{tomas.bzdusek@uzh.ch}
\affiliation{Department of Physics, University of Z\"urich, Winterthurerstrasse 190, 8057 Z\"urich, Switzerland}

\maketitle
\newpage

\onecolumngrid

\hspace*{0.1\textwidth}
\begin{minipage}{0.8\textwidth}
\begingroup
\hypersetup{linkcolor=black}
\tableofcontents
\endgroup
\end{minipage}
\clearpage

% Change figure labels to Supplementary Fig. 
\renewcommand{\figurename}{Supplementary Fig.}

\begin{figure}[p]
\centering
\includegraphics[width=1\linewidth]{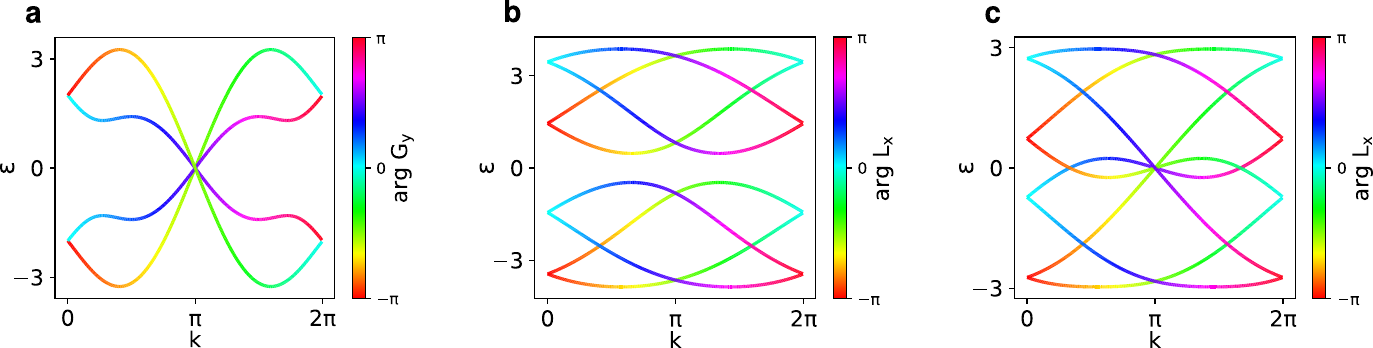}
\phantomsection
  \caption{
\textbf{Detailed band structures of the triangular ladder models.}  
\bfsf{a}~Bulk energy bands of the model in Fig.~2\bfsf{a} for $t_1 = t_2 = 1$, colored according to the phase of the eigenvalue of the glide operator $\mathsf{G}_y$ in Eq.~(9).
\bfsf{b,c} Bulk energy bands of the model in Fig.~2\bfsf{b} after Brillouin zone folding per Eq.~(21), colored according to the phase of the eigenvalue of the translation operator $\mathsf{L}_x$ 
in Eq.~(23), corresponding to parameter values $t_1 = 1,t_2 = 2$ (panel $\bfsf{b}$) and to $t_1 = t_2 =1$ (panel $\bfsf{c}$).
}
\label{fig:ladders-bands}
\figtocentry{Detailed band structures of the triangular ladder models.}
\end{figure}

\FloatBarrier

\twocolumngrid

\appsectitle{Explicit construction of CSL Hamiltonian}\label{app:regular-rep}

In this section, we consider explicit examples for the construction of the CSL Hamiltonian from Eq.~(2) of the main text. We first discuss the case $G = Z_2$, where we explicitly construct the regular representation matrices and demonstrate how different connections lead to distinct Hamiltonians distinguished by their Wilson loops. We end the discussion by sketching the construction for $G = Q_8$. 

\bigskip

We begin with considering the gauge group $Z_2$. Consider a pair of pillars, associated with sites $m$ and $n$, that are connected by a bond with hopping amplitude $t_{mn}$. 
Given that $G = Z_2 = \{+1, -1\}$, each pillar contains two orbitals, resulting in two basis states for both the pillar $m$ and for the pillar $n$:
\begin{equation}
\label{eqn:two-Z2s-basis}
|m, +1\rangle, \quad |m, -1\rangle ; \quad |n, +1\rangle, \quad |n, -1\rangle.
\end{equation}
Since the gauge group contains two elements, the choice of connection $a\in Z_2$ specifies one of two possible ways of coupling the orbitals at pillars $m$ and $n$.

\begin{figure}[b!]
  \centering  
  \includegraphics[width=0.5\linewidth]{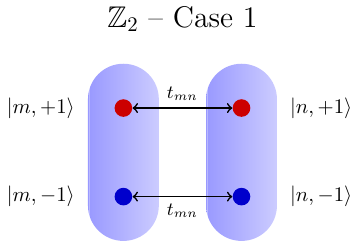}
  \caption{
  \textbf{Case 1: Trivial $\mathbf{Z_2}$ connection.}
  Two~pillars of a CSL with gauge group $Z_2$, related by connection $a = +1$ (``horizontal'' coupling). The two pillars correspond to sites $m$ and $n$. Each pillar (blue capsule) contains two orbitals, which are labeled by the group elements $\pm 1 \in Z_2$. 
  The connection $a = +1$ couples orbitals with matching labels: $|n, g\rangle \leftrightarrow |m, (+1)\cdot g\rangle = |m, g\rangle$. This is also shown in Fig.~1\bfsf{a} of the main text.}
  \label{fig:Z2_case1}
\end{figure}

\bigskip

In Case 1, the connection is $a = +1$  (``horizontal'' coupling). We consider two neighboring pillars $m$ and $n$ with connection $a_{m,n} = +1$. The CSL Hamiltonian for this bond is
\begin{equation}
H_{m,n} = t_{mn} \sum_{g,h \in G} c^\dagger_{m,g} \, [\rho(a_{m,n})]_{g,h} \, c_{n,h} + \mathrm{h.c.}
\end{equation}
where $\rho(a_{m,n})$ is the regular representation of the connection. The matrix elements are $[\rho(a)]_{g,h} = 1$ if $g = a \cdot h$ and zero otherwise. 

For $a_{m,n} = +1$, we construct $\rho(+1)$ in the basis $(+1, -1)$ by checking each matrix element:
\begin{align}
[\rho(+1)]_{+1,+1} &= 1 \quad \text{because} \quad +1 = (+1) \cdot (+1) \nonumber \\
[\rho(+1)]_{+1,-1} &= 0 \quad \text{because} \quad +1 \neq (+1) \cdot (-1) = -1 \nonumber \\
[\rho(+1)]_{-1,+1} &= 0 \quad \text{because} \quad -1 \neq (+1) \cdot (+1) = +1 \nonumber \\
[\rho(+1)]_{-1,-1} &= 1 \quad \text{because} \quad -1 = (+1) \cdot (-1)
\end{align}
This gives
\begin{equation}
\rho(+1) = \mathbb{1} = \begin{bmatrix} 1 & 0 \\ 0 & 1 \end{bmatrix}.
\label{eq:rho-horizontal}
\end{equation}
Orbitals with matching labels are coupled (Supplementary Fig.~\ref{fig:Z2_case1}): $|n, +1\rangle \leftrightarrow |m, +1\rangle$ and $|n, -1\rangle \leftrightarrow |m, -1\rangle$. 
In the basis of Eq.~(\ref{eqn:two-Z2s-basis}), this corresponds to the Hamiltonian
\begin{equation}
\label{eq:Ham-horizontal}
H_{m,n}^{a=+1} = t_{mn}\begin{bmatrix} \mathbb{0} & \rho(+1) \\ \rho(+1) & \mathbb{0} \end{bmatrix} = t_{mn}\begin{bmatrix} 0 & 0 & 1 & 0 \\ 0 & 0 & 0 & 1 \\ 1 & 0 & 0 & 0 \\ 0 & 1 & 0 & 0 \end{bmatrix},   
\end{equation}
where $\mathbb{0}$ is the zero $2\times 2$ matrix. 
This choice of connection reproduces the ``horizontal'' coupling in Fig.~1\bfsf{a} in the main~text.

\bigskip

\begin{figure}[b!]
  \centering  
  \includegraphics[width=0.5\linewidth]{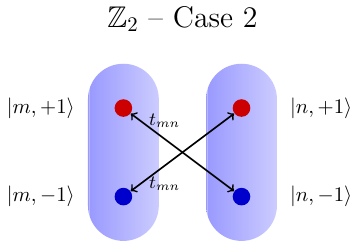}
  \caption{
  \textbf{Case 2: Non-trivial $\mathbf{Z_2}$ connection.}
  Two pillars of a CSL with gauge group $Z_2$, related by connection $a = -1$ (``swapped'' coupling). The connection $a = -1$ couples orbitals with opposite labels: $|n, g\rangle \leftrightarrow |m, (-1)\cdot g\rangle = |m, -g\rangle$. This is also shown in Fig.~1\bfsf{a} of the main text.}
  \label{fig:Z2_case2}
\end{figure}

In Case 2, the connection is $a = -1$ (``swapped'' coupling).    
For $a_{m,n} = -1$, we have $[\rho(-1)]_{g,h} = 1$ if $g = (-1) \cdot h$ and zero otherwise. Using $(-1) \cdot (+1) = -1$ and $(-1) \cdot (-1) = +1$, we get
\begin{equation}
\rho(-1) = \sigma_x = \begin{bmatrix} 0 & 1 \\ 1 & 0 \end{bmatrix}.
\label{eq:rho-swapped}
\end{equation}
Orbitals with opposite labels are coupled (Supplementary Fig.~\ref{fig:Z2_case2}): $|n, +1\rangle \leftrightarrow |m, -1\rangle$ and $|n, -1\rangle \leftrightarrow |m, +1\rangle$. 
In the basis of Eq.~(\ref{eqn:two-Z2s-basis}), this corresponds to the Hamiltonian
\begin{equation}
\label{eq:Ham-swapped}
H_{m,n}^{a=-1} = t_{mn}\begin{bmatrix} \mathbb{0} & \rho(-1) \\ \rho(-1) & \mathbb{0} \end{bmatrix} = t_{mn}\begin{bmatrix} 0 & 0 & 0 & 1 \\ 0 & 0 & 1 & 0 \\ 0 & 1 & 0 & 0 \\ 1 & 0 & 0 & 0 \end{bmatrix}.   
\end{equation}
This corresponds to the ``swapped'' coupling in Fig.~1\bfsf{a} in the main text.

\bigskip

While the Hamiltonians in Eqs.~(\ref{eq:Ham-horizontal}) and~(\ref{eq:Ham-swapped}) appear different, they have the same spectrum. 
This is because one can alter the connection $a=+1 \, \leftrightarrow \,a=-1$ by applying a nontrivial gauge transformation $\chi = -1$ to one of the two sites.

However, the two Hamiltonians become distinguished once we introduce a closed loop. 
For concreteness, let us interpret the two sites as a one-dimensional chain and introduce a periodic boundary condition (PBC). 
This introduces an additional outer coupling between the two sites, and we set the corresponding connection to $a_\textrm{PBC}=+1$ (the ``horizontal'' coupling). 
The Hamiltonian for the two choices of the connection $a$ on the inner bond of the chain becomes for $t_{mn} = t$
\begin{equation}
H^\textrm{\,Case 1}_\textrm{\,PBC} \!=\! t\begin{bmatrix}
\mathbb{0}\!\!\!  & \!\!\! \rho(+1) + \rho(1) \\
\rho(+1) + \rho(1) \!\!\! & \!\!\! \mathbb{0} 
\end{bmatrix} \!=\! t\begin{bmatrix}
0 & 0 & 2 & 0 \\
0 & 0 & 0 & 2 \\
2 & 0 & 0 & 0 \\
0 & 2 & 0 & 0
\end{bmatrix}
\end{equation}
and 
\begin{equation}
H^\textrm{\,Case 2}_\textrm{\,PBC} \! =  \! t\begin{bmatrix}
\mathbb{0}\!\!\!  & \!\!\! \rho(-1) + \rho(1) \\
\rho(-1) + \rho(1) \!\!\! & \!\!\! \mathbb{0} 
\end{bmatrix} \!=\! t\begin{bmatrix}
0 & 0 & 1 & 1 \\
0 & 0 & 1 & 1 \\
1 & 1 & 0 & 0 \\
1 & 1 & 0 & 0
\end{bmatrix}.
\end{equation}
One can verify that these two Hamiltonian have different spectra, implying that they cannot be related through a gauge transformation.
Specifically, they are distinguished by the Wilson loop $W_\textrm{chain} = a \cdot a_\textrm{PBC} = \pm 1$ along the one-dimensional chain, which is invariant under gauge transformations.

\bigskip

The Hamiltonian construction proceeds analogously for the gauge group $G = Q_8$. 
In this case, the pillar at each site contains eight orbitals, i.e., one for each element of $Q_8 = \{\pm 1, \pm\imi, \pm\imj, \pm\imk\}$.
Similarly, the choice of connection $a \in Q_8$ between any two sites encodes one of eight possible connectivities between the two octuplets of orbitals.

\begin{figure}[b!]
  \centering  
  \includegraphics[width=\linewidth]{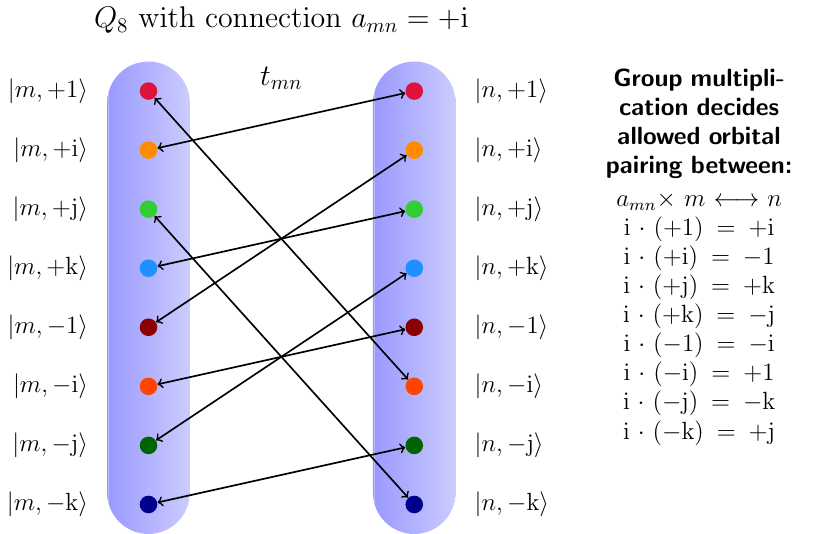}
  \caption{
  \textbf{Example of $\mathbf{Q_8}$ connection.}
  Two pillars of a CSL with gauge group $Q_8$, related by connection $a_{m,n} = +\imi$. 
  The right panel illustrates the allowed orbital pairings as dictated by the underlying group multiplication.}
  \label{fig:Q8_case}
\end{figure}

For concreteness, let us consider two sites $n$ and $m$ with connection $a_{m,n} = +\imi$ (Supplementary Fig.~\ref{fig:Q8_case}).
In this case, the Hilbert space is spanned by sixteen orbitals, namely 
\begin{equation}
\label{eqn:two-Q8s-basis}
\begin{split}
|m, +1\rangle, \quad |m, +\imi\rangle, \quad |m, +\imj\rangle, \quad |m, +\imk\rangle , \\
|m, -1\rangle, \quad |m, -\imi\rangle, \quad |m, -\imj\rangle, \quad |m, -\imk\rangle , \\
|n, +1\rangle, \quad |n, +\imi\rangle, \quad |n, +\imj\rangle, \quad |n, +\imk\rangle , \\
|n, -1\rangle, \quad |n, -\imi\rangle, \quad |n, -\imj\rangle, \quad |n, -\imk\rangle . 
\end{split}
\end{equation}
The basis state $|m,g\rangle$ is connected to the basis state $|n,h\rangle$ by a hopping term $t_{mn}$ if and only if $g = a_{m,n} \cdot h$. 
This connectivity is encoded by the regular representation $\rho(a_{m,n})$, which is an $8 \times 8$ permutation matrix such that $[\rho(a_{m,n})]_{g,h} = 1$ if $g = a_{m,n} \cdot h$ and zero otherwise. 
Following the products displayed in the right side of Supplementary Fig.~\ref{fig:Q8_case} (and assuming the ordering of the eight elements as $\{+1,+\imi,+\imj,+\imk,-1,-\imi,-\imj,-\imk\}$), we find the explicit form of this matrix as
\begin{equation}
\rho(+\mathrm{i}) = \begin{bmatrix}
0 & 0 & 0 & 0 & 0 & 1 & 0 & 0 \\
1 & 0 & 0 & 0 & 0 & 0 & 0 & 0 \\
0 & 0 & 0 & 0 & 0 & 0 & 0 & 1 \\
0 & 0 & 1 & 0 & 0 & 0 & 0 & 0 \\
0 & 1 & 0 & 0 & 0 & 0 & 0 & 0 \\
0 & 0 & 0 & 0 & 1 & 0 & 0 & 0 \\
0 & 0 & 0 & 1 & 0 & 0 & 0 & 0 \\
0 & 0 & 0 & 0 & 0 & 0 & 1 & 0
\end{bmatrix},
\end{equation}
and the $16\times 16$ Hamiltonian matrix with the basis in Eq.~(\ref{eqn:two-Q8s-basis}) takes the form
\begin{equation}
\label{eqn:Ham-Q8-two-pillars}
H_{m,n} = t_{mn}\begin{bmatrix}
\mathbb{0} & \rho(+\imi) \\
\rho(+\imi)^{-1} & \mathbb{0}
\end{bmatrix}.
\end{equation}
Note that $\rho(g)^{-1} = \rho(g^{-1})$ because $\rho$ is a representation of the gauge group, ensuring Hermiticity of the Hamiltonian.

The full CSL Hamiltonian, defined by Eq.~(2) in the main text, involves the summation of Hamiltonian blocks analogous to Eq.~(\ref{eqn:Ham-Q8-two-pillars}), where the regular representation $\rho(a_{m,n})$ on the bond from $n$ to $m$ takes one of eight possible forms (one for each element in $Q_8$).

\appsectitle{Automorphisms of the quaternion group}\label{app:aut-group}

Here we provide an elucidation of the automorphism group of the quaternion group.
Further discussion that reveals why composition of a local gauge transformation with a global gauge automorphism encompasses the most general transformation of CSL models is included in Sec.~\ref{app:aut-only-globally} of the SIF.

Since two automorphisms $\varphi_1$ and $\varphi_2$ of a group $G$ can be composed, $\textrm{Aut}(G)$ is itself a group, called the automorphism group of $G$. 
All automorphisms preserve the conjugacy class $\{+1\}$ of the identity element; in addition, they can only relate conjugacy classes containing the same number of elements, implying that $\varphi(-1) = -1$ is also invariant.
It follows that there are $24$ automorphisms of $Q_8$, corresponding to the six permutations of conjugacy classes $\{\pm \imi\}$, $\{\pm \imj\}$, $\{\pm \imk\}$ times the four signs of $\varphi(+\imi)$ and $\varphi(+\imj)$ [with the third sign fixed by $\varphi(+\imi)\cdot \varphi(+\imj)=\varphi(+\imk)$].
A careful analysis reveals that $\textrm{Aut}(Q_8) = S_4$ (the permutation group of four elements).

To better understand the structure of the automorphism group $\textrm{Aut}(Q_8)$, recall that, for a general choice of $G$, $\textrm{Aut}(G)$ can be decomposed into two special components: inner automorphism $\textrm{Inn}(G)$ and outer automorphisms $\textrm{Out}(G)$.
Mathematically, this decomposition works in the sense that the short sequence of homomorphisms
\begin{equation}
1 \to \textrm{Inn}(G) \to \textrm{Aut}(G) \to \textrm{Out}(G) \to 1
\end{equation} 
is exact, i.e., $\textrm{Aut}(G)$ is an extension of $\textrm{Out}(G)$ by $\textrm{Inn}(G)$. 
In most practical examples (including the quaternion group $Q_8$), this extension amounts to a semidirect product of the two components.

Specifically, elements $\varphi^{(h)} \in \textrm{Inn}(G)$ correspond to conjugation of elements in $G$ with a fixed element $h \in G$, i.e., 
\begin{equation}
\label{eqn:inner-auto-interpret}
\varphi^{(h)}: g \mapsto h \cdot g \cdot h^{-1}.
\end{equation}
For the quaternion group, such conjugations can at best result in a sign flip of the imaginary units; therefore, $\textrm{Inn}(Q_8) = Z_2 \times Z_2$ is the Klein four-group, with the two factors interpreted as the signs of $\varphi(+\imi)$ and $\varphi(+\imj)$.
However, the interpretation of inner automorphisms per Eq.~(\ref{eqn:inner-auto-interpret}) implies that, within the CSL framework, they are equivalent to performing a global gauge transformation with $\chi_i = h$. 
Therefore, $\textrm{Inn}(G)$ does not extend the theoretical framework for representing space group symmetries beyond projective representations.

The second (i.e., outer) component of $\textrm{Aut}(G)$ is obtained by observing that $\textrm{Inn}(G)$ constitutes a normal subgroup; therefore, it is possible to define the quotient $\textrm{Aut}(G)/\textrm{Inn}(G) = \textrm{Out}(G)$.
For the quaternion group, a known result $\textrm{Out}(Q_8) = S_3$ corresponds to the permutation of the conjugacy classes of the imaginary units.
The adjective `outer' suggests that each of these further automorphisms can be understood as a conjugation with some element outside of $G$. 
Specifically for $Q_8$, assuming its embedding inside $SU(2)$ according to Eq.~(7), every $\varphi \in \textrm{Out}(G)$ can be interpreted as a conjugation with certain unitaries in the complement $SU(2)\backslash Q_8$.
The involvement of transformations outside the gauge group goes beyond the usual formalism of projectively represented symmetries. 
We refer to symmetries that require a composition with an outer automorphism as being $\textrm{Aut}(G)$-modified.
We leave the systematic characterization of $\textrm{Aut}(G)$-modified symmetries to future studies.

\appsectitle{Transformations preserving the gauge structure}
\label{app:aut-only-globally}

\begin{figure}[b!]
  \centering  
  \includegraphics[width=\linewidth]{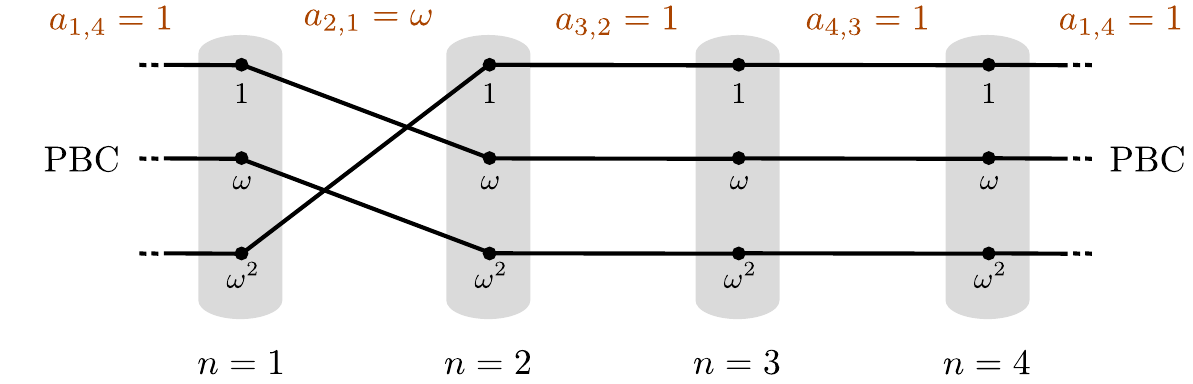}
  \caption{
  \textbf{Understanding local gauge transformations (part I).}
  CSL with gauge group $G=Z_3$ and nearest-neighbor coupling over a one-dimensional chain with four sites labeled by $n\in\{1,\ldots,4\}$. The connection $a_{n+1,n}$ is indicated with brown font over the bonds and `PBC' stands for `periodic boundary condition'. 
  Observe that the system is set up so that the twelve constituent orbitals are arranged into a single loop of length $L=12$.}
  \label{fig:Z3-CSL-fig1}
\end{figure}

It may appear mysterious why we allow for local gauge transformations (as inherent in any gauge theory) while limiting our attention to only global (i.e., not spatially varying) automorphisms of the gauge group. 
In this section, we address this aspect of CSLs as introduced in the main text.
Our discussion is structured as follows. First, we consider a particularly simple example to provide intuition into gauge transformations and automorphisms of the gauge group, showcasing explicitly how application of a local gauge automorphism leads to a contradiction in the mathematical description. 
(To simplify the terminology, we adopt in the following the shorter phrase gauge automorphism to mean automorphism of the gauge group.)
Second, we use rigorous mathematical reasoning to derive that the most general transformation preserving the gauge structure of CSL models corresponds to a composition of a local gauge transformation with a global gauge automorphism.

To develop intuition into the role of gauge transformations and gauge automorphisms, we simplify the discussion by considering the cyclic group $Z_3 = \{1,\omega,\omega^2\}$ as the gauge group $G$ instead of the quaternion group. 
The discussion of $G=Q_8$ follows the same logic while involving a larger number of orbitals.
It can be shown that $\textrm{Aut}(Z_3) = Z_2$, with the non-trivial element $\varphi$ swapping $\omega \leftrightarrow \omega^2$.

\bigskip

Recall that a CSL constructed over $N$ sites with gauge group $G$ involves $N \times \abs{G}$ constituent orbitals. 
The CSL construction neatly arranges these orbitals into $N$ pillars, each pillar containing $\abs{G}$ orbitals. 
Nevertheless, one is free to treat the CSL as a tight-binding model over $N \times \abs{G}$ sites (even if this perspective loses access to some structure). 
In the following paragraphs, we temporarily adopt such a relaxed perspective to help us track the modifications made to the model under gauge transformations and gauge automorphisms. 
For concreteness, consider a periodic chain of $N=4$ sites labeled with $n\in\{1,2,3,4\}$ with $\abs{Z_3} = 3$ orbitals per site. 
We assume nearest-neighbor hopping amplitude $t$, and we set $a_{2,1} = \omega$ and $a_{3,2} = a_{4,3} = a_{1,4} = 1$ as shown in Supplementary Fig.~\ref{fig:Z3-CSL-fig1}.

\begin{figure}[b!]
  \centering  
  \includegraphics[width=\linewidth]{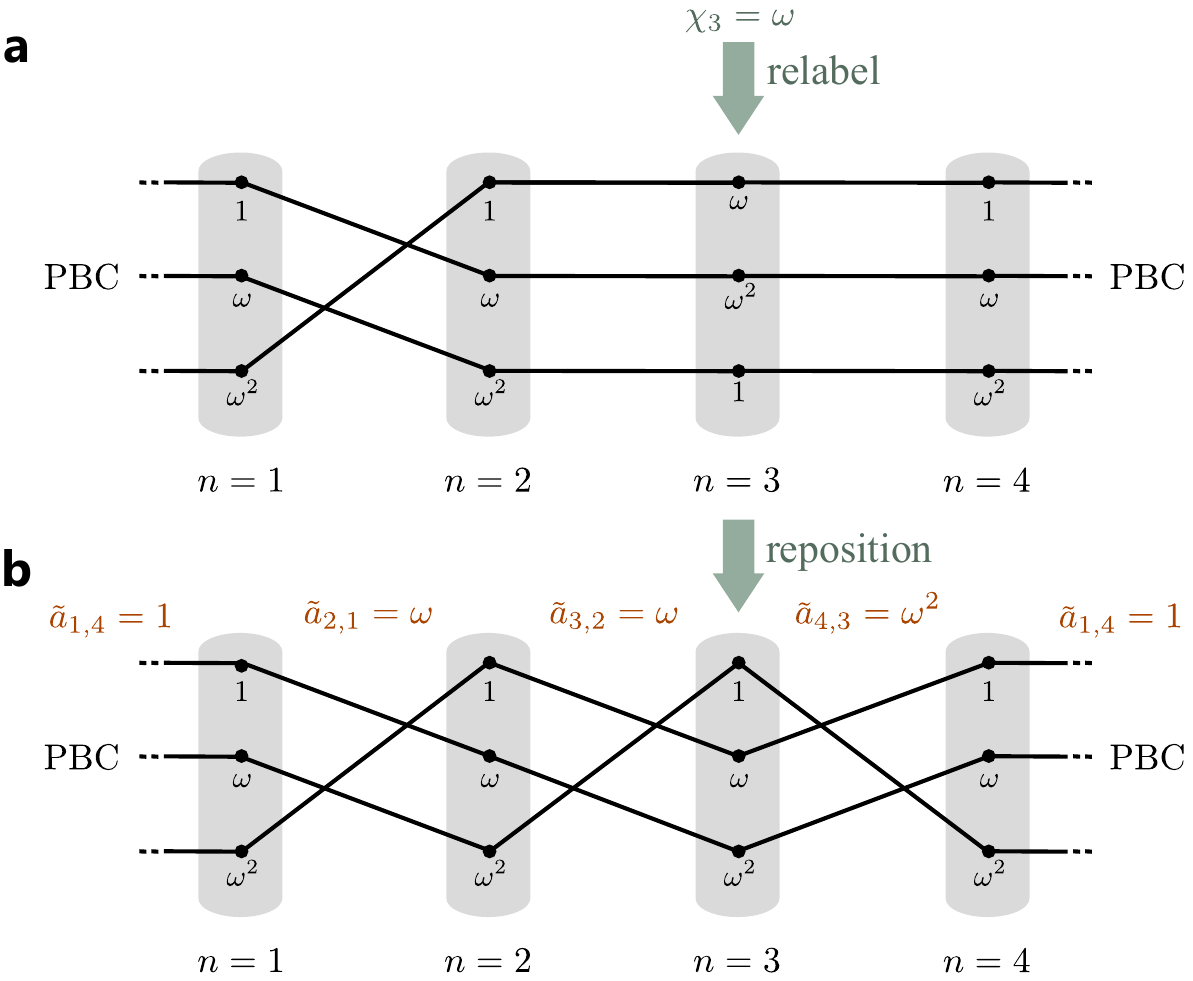}
  \caption{
  \textbf{Understanding local gauge transformations (part II).}
  \bfsf{a}~Starting with the situation in Supplementary Fig.~\ref{fig:Z3-CSL-fig1}, local gauge trans\-for\-ma\-tion with $\chi_3  = \omega$ relabels the orbitals at site $n=3$ through $g \mapsto \omega \cdot g$ while preserving the connectivity. 
  \bfsf{b}~Repositioning the orbitals within the third pillar according to their labels $g\in G$ into a canonical ordering (while preserving connectivity), we recognize that the connection at the adjacent bonds has been changed to $\tilde{a}_{3,2} = \omega$ and $\tilde{a}_{4,3} = \omega^2$.
  The system is unchanged (still exhibiting the closed loop of $L=12$ sites) and the connection remains well-defined (valued in $G=Z_3$ on all bonds).}
  \label{fig:Z3-CSL-fig2}
\end{figure}

We first consider the effect of a local gauge transformation, which corresponds to a local relabeling of the sites while keeping the overall connectivity of (and therefore the tight-binding model over) the $N\times \abs{G}$ constituent orbitals invariant. 
To showcase this explicitly, we apply a non-trivial $\chi_3 = \omega$ to site $n=3$ (with trivial $\chi_1 = \chi_2 = \chi_4 = 1$ at the other sites) to the model in Supplementary Fig.~\ref{fig:Z3-CSL-fig1}. 
The gauge transformation changes the connection on the adjacent bonds [by virtue of Eq.~(3) in the Results section] to $\tilde{a}_{3,2} = \omega$ and $\tilde{a}_{4,3} = \omega^2$.
Despite these changes, we recognize from Supplementary Fig.~\ref{fig:Z3-CSL-fig2}\bfsf{b} that the graph over the constituent orbitals remains unchanged, i.e., twelve cyclically connected sites. 
Crucially, the connection $\tilde{a}_{n+1,n}$ on each bond remains well defined.

Next, consider the application of the non-trivial global automorphism $\varphi: \omega \leftrightarrow \omega^2$ to the model in Supplementary Fig.~\ref{fig:Z3-CSL-fig2}\bfsf{b}. 
This corresponds to globally relabeling the $\omega \leftrightarrow \omega^2$ elements of all the pillars as shown in Supplementary Fig.~\ref{fig:Z3-CSL-fig3}\bfsf{a}. 
In the second step, illustrated in Supplementary Fig.~\ref{fig:Z3-CSL-fig3}\bfsf{b}, we move the orbitals within each pillar to their standard ordering $\{1,\omega,\omega^2\}$ while preserving their connectivity. 
We observe that the connection has changed in a manner consistent with the gauge automorphism: the initial connection was $\tilde{a}_{n+1,n} = \omega$; the new connection has become $\hat{a}_{n+1,n} = \omega^2$ (and vice versa), while elements equal to the identity $1$ have remained unchanged. 
Also, one can track that the system still corresponds to a loop of cyclically coupled twelve orbitals.
In summary: the global application of the non-trivial gauge automorphism keeps the system invariant; it is only our description of the system that has been altered (as anticipated of a gauge theory). 
Furthermore, the connection remains well-defined, taking values $\hat{a}_{n+1,n} \in Z_3$ for all bonds.

\begin{figure}[t!]
  \centering  
  \includegraphics[width=\linewidth]{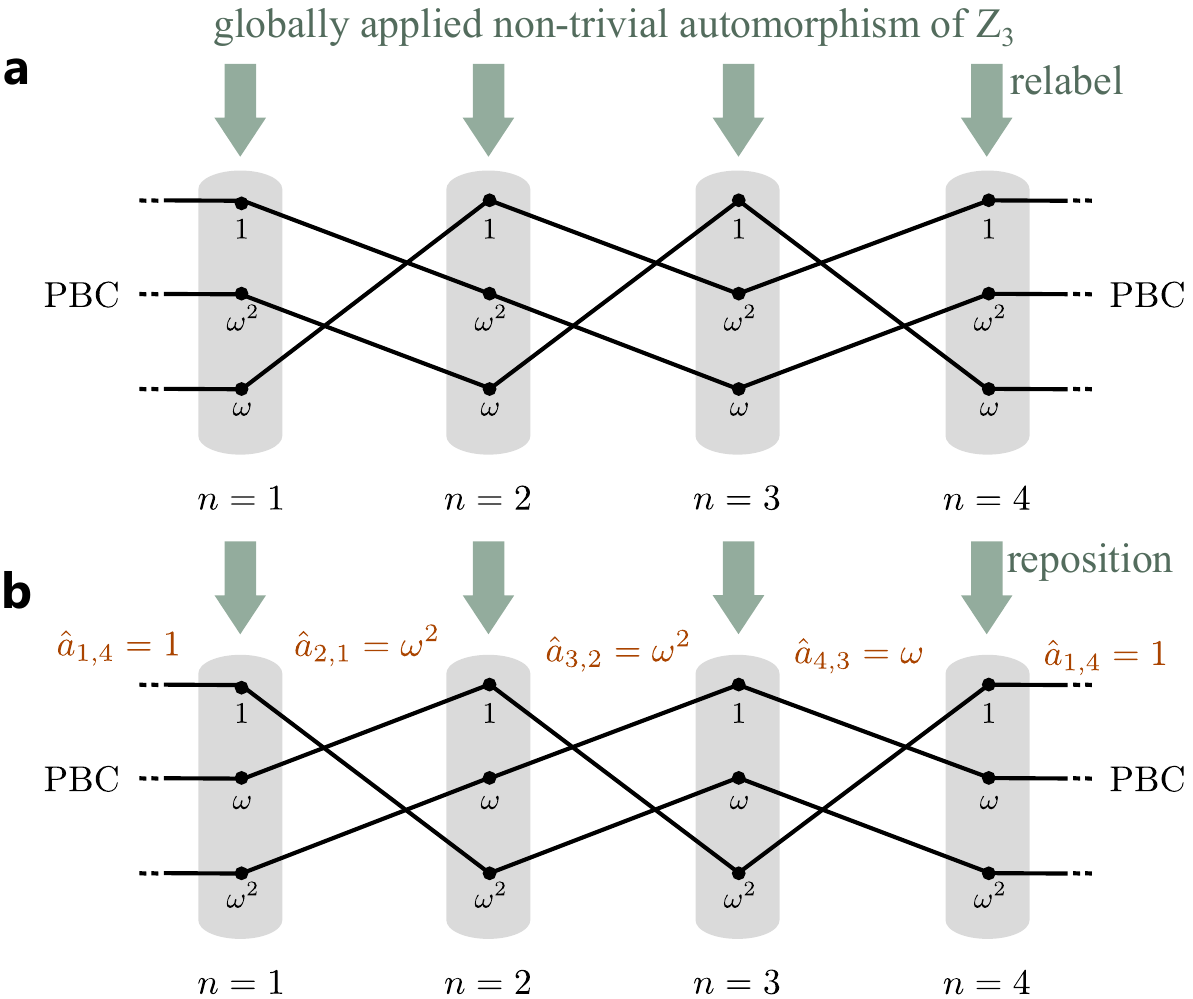}
  \caption{
  \textbf{Understanding global automorphisms (part I).}
  \bfsf{a}~Starting with the situation in Supplementary Fig.~\ref{fig:Z3-CSL-fig2}\bfsf{b}, global gauge auto\-mor\-phism $\varphi$ relabels the orbitals $\omega \leftrightarrow \omega^2$ at all sites. 
  \bfsf{b}~We reposition the orbitals within the individual pillars into the standard ordering $\{1,\omega,\omega^2\}$ while preserving their connectivity.
  We recognize that the coupling of adjacent pillars is still captured with well-defined connection $\hat{a}_{n+1,n} \in Z_3$; namely, $\hat{a}_{n+1,n} = \varphi(\tilde{a}_{n+1,n})$.
  The system is unchanged (still exhibiting the closed loop of $L=12$ sites), meaning that it is only our description of that system that has been altered.}
  \label{fig:Z3-CSL-fig3}
\end{figure}

We contrast the previous consideration with the local application of the non-trivial automorphism $\varphi$. 
For concreteness, we apply $\varphi$ to site $n=2$ of the model in Supplementary Fig.~\ref{fig:Z3-CSL-fig1}. 
This corresponds to relabeling the $\omega \leftrightarrow \omega^2$ orbitals in the second pillar as shown in Supplementary Fig.~\ref{fig:Z3-CSL-fig4}\bfsf{a}.
We then reposition the orbitals to the standard ordering $\{1,\omega ,\omega^2\}$ while preserving the connectivity, leading to the situation illustrated in Supplementary Fig.~\ref{fig:Z3-CSL-fig4}\bfsf{b}.
We find that the coupling of the orbitals in pillar $n=2$ to the adjacent pillars $n\in\{1,3\}$ does not follow multiplication with a connection valued in $Z_3$. 
Indeed, focusing on the adjacent pillars $n\in\{1,2\}$, there is no connection $a'_{2,1} \in Z_3$ that simultaneously solves
\begin{equation}
a'_{2,1} \cdot 1 = \omega^2, \quad
a'_{2,1} \cdot \omega = \omega \quad \textrm{and} \quad
a'_{2,1} \cdot \omega^2 = 1.
\end{equation}
This contradiction exemplifies why applications of local gauge automorphisms are excluded from the gauge freedom of CSLs.

\begin{figure}[t!]
  \centering  
  \includegraphics[width=\linewidth]{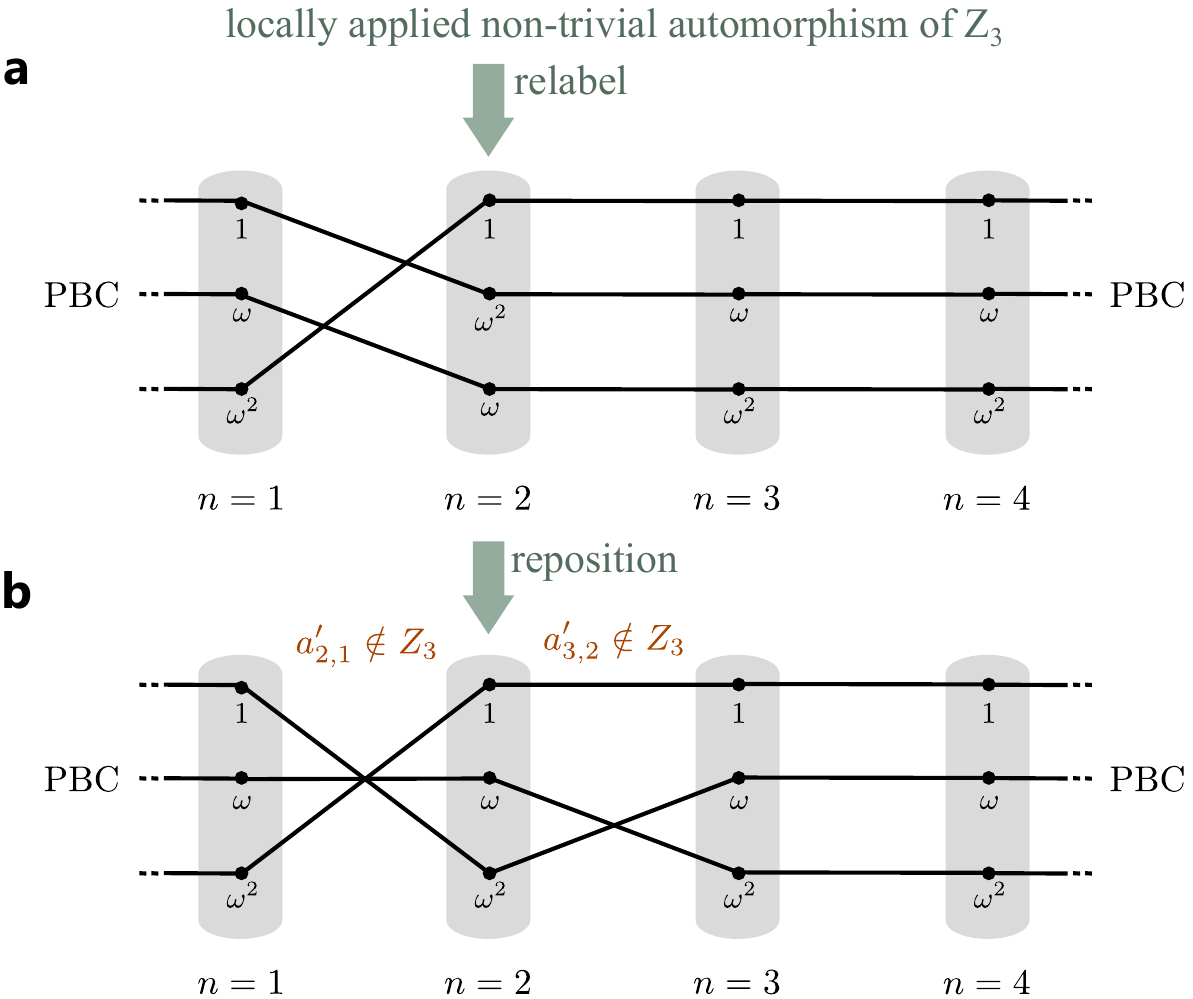}
  \caption{
  \textbf{Understanding global automorphisms (part II).}
  \bfsf{a}~Starting with the situation in Supplementary Fig.~\ref{fig:Z3-CSL-fig1}, local gauge auto\-mor\-phism $\varphi$ relabels the orbitals $\omega \leftrightarrow \omega^2$ at site $n=2$. 
  \bfsf{b}~We reposition the orbitals within the second pillar into the standard ordering $\{1,\omega,\omega^2\}$ while preserving their connectivity.
  Observe that the coupling of the orbitals in the first and in the second pillar (and similarly for the second and third pillar) does not correspond to multiplication with a connection $a'_{2,1}$ valued in $Z_3$ (and similarly for $a'_{3,2}$).
  While the system remains invariant (namely, a loop of twelve cyclically coupled sites), the local gauge automorphism brought us out of the gauge description with $G=Z_3$.}
  \label{fig:Z3-CSL-fig4}
\end{figure}

Let us remark that the contradiction arises even though the system in Supplementary Fig.~\ref{fig:Z3-CSL-fig4}\bfsf{b} is still equivalent to the original one, i.e., an arrangement of twelve sites that are cyclically coupled. 
In summary: a local application of the gauge automorphism keeps the system invariant; however, it breaks down the description in terms of a well-defined connection $a'_{n+1,n} \in Z_3$. Since such a process takes us outside of the specified mathematical playground of CSLs, we show that only global gauge automorphisms can be applied to represent symmetry operators of the corresponding models.

\bigskip

Motivated by the explicit example discussed above, we put the notion of global gauge automorphisms on a mathematically solid footing for a general choice of gauge group $G$ and for an arbitrary CSL model.
Let us consider two adjacent sites $n$ and $\ell$ coupled with connection $a_{n,\ell}$, i.e., the orbital $g_\ell$ at site $\ell$ (state $\ket{\ell,g_\ell}$ of the Hilbert space) is coupled to the orbital $g_n = a_{n,\ell}\cdot g_\ell$ at site $n$ (state $\ket{n,a_{n,\ell}\cdot g_\ell}$ of the Hilbert space).
Let us consider the most general permutations $\sigma_{n}$ and $\sigma_{\ell}$ of the orbital labels at sites $n$ and $\ell$.
We aim to determine the most general choice of $\sigma_n$ and $\sigma_\ell$ such that, after the relabeling, the coupling of pillars $n$ and $\ell$ corresponds to a connection $\tilde{a}_{n,\ell}$ valued in $G$.

We readily know that two types of solutions exist. 
First, $\sigma_{n}$ and $\sigma_\ell$ can be local gauge transformations, i.e., left multiplication with $\chi_n$ at site $n$ and with $\chi_\ell$ at site $\ell$; then, the connection is altered to $\tilde{a}_{n,\ell} = \chi_n \cdot a_{n,\ell} \cdot \chi_\ell^{-1}$ [Eq.~(3) in the Results section].
Second, we can apply a global automorphism $\varphi$; through the defining aspect of automorphism that $\varphi(x\cdot y) = \varphi(x)\cdot \varphi(y)$ for any $x,y\in G$, we find that the orbital $\varphi(g_\ell)$ at site $\ell$ is connected to $\varphi(a_{n,\ell}\cdot g_\ell) = \varphi(a_{n,\ell})\cdot \varphi(g_\ell)$ at site $n$, i.e., the new connection is $\tilde{a}_{n,\ell} = \varphi(a_{n,\ell})$. 
We claim that the composition of a local gauge transformation with a global gauge automorphism constitutes the only solutions to the specified problem. 
We prove this statement~below.

The original bond couples orbital $\ket{\ell, g_\ell}$ to orbital $\ket{n,a_{n,\ell} \cdot g_\ell}$. 
This statement applies for any choice of $g_\ell$; therefore, we will drop the subscript in the following discussion. 
After the relabeling with $\sigma_{n}$ and $\sigma_\ell$, the same bond couples the orbital $\ket{\ell,\sigma_\ell(g)}$ to the orbital $\ket{n,\sigma_n(a_{n,\ell} \cdot g)}$.
For this coupling to be describable by a connection $\tilde{a}_{n,\ell}\in G$, we must have
\begin{equation}
\label{eqn:consistency-criterion}
\forall g \in G \; : \;  \sigma_n(a_{n,\ell}\cdot g) = \tilde{a}_{n,\ell} \cdot \sigma_\ell(g). 
\end{equation}
To proceed, we extract the group elements $\chi_n := \sigma_n(1)$ and $\chi_\ell := \sigma_\ell(1)$ (which will play the role of the local gauge transformation), and we use them to define new permutations $\pi_n(g) = \chi_n^{-1} \cdot \sigma_n(g)$ at site $n$ as well as $\pi_\ell(g) = \chi_\ell^{-1} \cdot  \sigma_\ell(g)$ at site $\ell$. 
Equivalently, we can express the initial permutations $\sigma$ through the newly introduced permutations $\pi$ as
\begin{equation}
\sigma_n(g) = \chi_n \cdot \pi_n(g) \quad \textrm{and} \quad \sigma_\ell(g) = \chi_\ell \cdot \pi_\ell(g).
\label{eqn:ungauged-permutation}
\end{equation}
Such a substitution ensures that
\begin{equation}
\label{eqn:reduced-permutations-to-1}
\pi_n(1) = \pi_\ell(1) = 1.    
\end{equation}
Substituting Eq.~(\ref{eqn:ungauged-permutation}) to Eq.~(\ref{eqn:consistency-criterion}) gives
\begin{equation}
\forall g \in G \; : \; \chi_n \cdot \pi_n(a_{n,\ell} \cdot g) = \tilde{a}_{n,\ell} \cdot \chi_\ell \cdot \pi_\ell(g).
\end{equation}
Multiplying from the left by $\chi_n^{-1}$ and introducing $b_{n,\ell} = \chi_n^{-1} \cdot \tilde{a}_{n,\ell} \cdot \chi_\ell \in G$, we obtain
\begin{equation}
\label{eqn:consistency-after-substitution}
\pi_{n}(a_{n,\ell}\cdot g) = b_{n,\ell} \cdot \pi_\ell(g) \quad\textrm{with $b\in G$}.
\end{equation}   
Evaluating Eq.~(\ref{eqn:consistency-after-substitution}) at $g = 1$ shows that
\begin{equation}
\pi_n(a_{n,\ell}) = b_{n,\ell} \cdot \pi_\ell(1) = b_{n,\ell},    
\end{equation}
which upon substitution back to Eq.~(\ref{eqn:consistency-after-substitution}) gives
\begin{equation}
\label{eqn:reduced-permutations-condition}
\pi_n(a_{n,\ell}\cdot g) = \pi_n(a_{n,\ell}) \cdot \pi_\ell(g).   
\end{equation}
Thus, after stripping away the left multiplications by $\chi_n,\chi_\ell$ according to Eq.~(\ref{eqn:ungauged-permutation}), we are left with the reduced permutations $\pi_n,\pi_\ell$ that satisfy Eqs.~(\ref{eqn:reduced-permutations-to-1}) and~(\ref{eqn:reduced-permutations-condition}). 
Observe that Eq.~(\ref{eqn:reduced-permutations-condition}) is almost the condition for an automorphism of a group except that two permutations on the right-hand side are labeled by different subscripts $n$ and $\ell$ respectively.

To finalize the proof, note that the presented construction must be possible for any choice of connection $a_{n,\ell}$.
Therefore, we actually have
\begin{equation}
\label{eqn:constraint-for-aut}
\forall \, a_{n,\ell},g \in G \; : \; \pi_n(a_{n,\ell}\cdot g) = \pi_n(a_{n,\ell}) \cdot \pi_\ell(g).  
\end{equation}
We claim that this is possible only if $\pi_n = \pi_\ell$. This is shown easily by setting $a_{n,\ell} =1$ in Eq.~(\ref{eqn:constraint-for-aut}); namely, 
\begin{eqnarray}
\pi_n(g) &=& \pi_n(1\cdot g) \stackrel{\textrm{(\ref{eqn:constraint-for-aut})}}{=} \pi_n(1) \cdot \pi_\ell(g) \nonumber \\
&\stackrel{\textrm{(\ref{eqn:reduced-permutations-to-1})}}{=}& 1 \cdot \pi_\ell(g) = \pi_\ell(g).   
\end{eqnarray}
Therefore, the reduced permutations $\pi_n = \pi_\ell \equiv \pi$ must be identical for any two connected sites (and, by extension, in the entire lattice).
Eq.~(\ref{eqn:constraint-for-aut}) thus becomes
\begin{equation}
\forall \, a,g \in G \; : \; \pi(a\cdot g) = \pi(a) \cdot \pi(g),     
\end{equation}
which constrains the global permutation $\pi: G \to G$ to be an automorphism of $G$.

Returning to Eq.~(\ref{eqn:ungauged-permutation}), we find that the most general transformation that preserves the gauge structure with connection valued in $G$ corresponds to permuting the orbitals in pillar $n$ with 
\begin{equation}
\sigma_n(g) = \chi_n \cdot \pi(g)    
\end{equation}
where $\chi_n = \sigma_n(1)$ is a local gauge transformation at site $n$ and $\pi$ is a global gauge automorphism. 
This general transformation replaces the original connection $a_{n,\ell}$ by
\begin{equation}
\tilde{a}_{n,\ell} = \chi_n \cdot \pi(a_{n,\ell}) \cdot \chi_\ell^{-1},
\end{equation}
which corresponds to the generalized version of Eq.~(3) in the Results section.

\appsectitle{Achieving energy separation of the irrep sectors}\label{App:pillar-Ham}

The Peter-Weyl decomposition in Eq.~(6) separates the dynamics of the CSL into blocks transforming in different irreps of the gauge group.
Unless additional steps are considered, the energy bands of each sector are centered at the same energy (at $\varepsilon=0$ for our models), implying that their energy bands overlap.
While this is not an obstacle for the experimental implementation as discussed in the context of electric circuits in the Results section, we show how the CSL construction can be further extended to achieve arbitrary splitting of the Hamiltonian sectors.

Our strategy is to furnish each site with a local `pillar-Hamiltonian' $H_{G\updownarrow}$, which is invariant under gauge transformations with the elements in $G$.
For simplicity, we first clarify this strategy for the case $G=Z_2$. 
Observing that the non-trivial gauge transformation ($\chi_n = -1$) permutes the two orbitals within a pillar with regular representation $\rho(-1) = \sigma_x$, the condition on invariance of the pillar Hamiltonian translates to $H_{Z_2 \updownarrow} = \sigma_x H_{Z_2\updownarrow} \sigma_x$.
We recognize the most general form as $H_{Z_2 \updownarrow} = \epsilon_0 \mathbb{1} + w \sigma_x$ with real coefficients $\epsilon_0$ and $w$. 
After diagonalization with the unitary prescribed by Eq.~(11), we find 
\begin{equation}
U^\dagger H_{Z_2 \updownarrow} U = \begin{bmatrix}
        \epsilon_0 + w & 0 \\ 0 & \epsilon_0 - w 
    \end{bmatrix},
\end{equation}
meaning that sectors transforming in the trivial (non-trivial) irrep of $Z_2$ exhibit an overall energy shift by $\varepsilon_0 + w$ (by $\varepsilon_0 - w$).
Such shifts are naturally realized in acoustic resonators, such as those of Ref.~\citenum{Xue:2022}, where the symmetric and the anti-symmetric cavity mode naturally occur at different energies. 

This construction naturally generalizes to other choices of the gauge group $G$. 
To find the most general form of $H_{G \updownarrow}$, it is sufficient to consider the equation 
\begin{equation}
\label{eqn:pillar-Ham-invariance}
H_{G \updownarrow} = \rho(g)^\dagger H_{G \updownarrow} \rho(g)    
\end{equation}
with $g$ the generators of $G$ (and with $\rho(g)$ their regular representation).
For the quaternion group, this amounts to taking $g= +\imi$ and $g=+\imj$. 
Assuming the ordering of the eight elements as $(+1, +\imi, +\imj,+\imk,-1,-\imi,-\imj,-\imk)$, we find
\begin{equation}
\label{eqn:Q8-pillar-Ham}
H_{Q_8 \updownarrow}=
\begin{bmatrix}
 \epsilon_0 & d_x & d_y & d_z & w & d_x^* & d_y^* & d_z^* \\
 d_x^* & \epsilon_0 & d_z^* & d_y & d_x & w & d_z & d_y^* \\
 d_y^* & d_z & \epsilon_0 & d_x^* & d_y & d_z^* & w & d_x \\
 d_z^* & d_y^* & d_x & \epsilon_0 & d_z & d_y & d_x^* & w \\
 w & d_x^* & d_y^* & d_z^* & \epsilon_0 & d_x & d_y & d_z \\
 d_x & w & d_z & d_y^* & d_x^* & \epsilon_0 & d_z^* & d_y \\
 d_y & d_z^* & w & d_x & d_y^* & d_z & \epsilon_0 & d_x^* \\
 d_z & d_y & d_x^* & w & d_z^* & d_y^* & d_x & \epsilon_0 \\
\end{bmatrix}
\end{equation}
where $\epsilon_0$ and $w$ are real and $d_{x,y,z}$ are complex coefficients.
Rotation of the basis with the unitary in Eq.~(11) results in $
U^\dagger H_{Q_8\updownarrow} U = \bigoplus_{K} H_\updownarrow^{K}$ with energies
\begin{subequations}
\label{eqn:1D-irrep-energy-shifts}
\begin{eqnarray}
H_\updownarrow^{A_0} &=& \epsilon_0 + w + 2\Re[d_{x} + d_{y} + d_{z}] \\ 
H_\updownarrow^{A_x} &=& \epsilon_0 + w + 2\Re[d_{x} - d_{y} - d_{z}] \\ 
H_\updownarrow^{A_y} &=& \epsilon_0 + w + 2\Re[-d_{x} + d_{y} - d_{z}] \\ 
H_\updownarrow^{A_z} &=& \epsilon_0 + w + 2\Re[-d_{x} - d_{y} + d_{z}]
\end{eqnarray}
\end{subequations}
for the one-dimensional irreps and with the $4 \times 4$ block
\begin{equation}
\label{eqn:pillar-Ham-2D-blocks}
H_\updownarrow^{E} = \Big[(\epsilon_0 -w)\mathbb{1}_\sigma - 2 \sum_j \Im[d_j]\sigma_\ell\Big] \otimes \mathbf{1}_\tau
\end{equation}
for the two copies of the two-dimensional irrep. 
In Eq.~(\ref{eqn:pillar-Ham-2D-blocks}), the Pauli matrices $\tau_j$ act on the two orbitals within a chosen copy of the 2D irrep while the Pauli matrices $\sigma_j$ mix the two copies of the 2D irrep.

The expression in Eq.~(\ref{eqn:pillar-Ham-2D-blocks}) can be transformed to $\left[(\epsilon_0 -w)\mathbf{1}_\sigma - 2 d_\textrm{I}  \sigma_z \right] \otimes \mathbb{1}_\tau$ with $d_\textrm{I} = (\sum_{\ell} \Im[d_\ell]^2)$ by utilizing a suitable unitary rotation of the matrices $D^E(g)$ in Eq.~(7). 
Therefore, $H_{Q_8 \updownarrow}$ shifts the band structures of the individual sectors by
\begin{subequations}
\begin{eqnarray}
\epsilon^{A_0} &=& \epsilon_0 + w + 2 \textstyle{\sum_j} \Re[d_j] \\    
\epsilon^{A_\ell} &=& \epsilon_0 + w - 2 \textstyle{\sum_j} \Re[d_j] + 4 \Re[d_\ell]\;\;\;\; \\
\epsilon^{(E,1)} &=& \epsilon_0 - w - 2d_\textrm{I} \\
\epsilon^{(E,2)} &=& \epsilon_0 - w + 2d_\textrm{I}.
\end{eqnarray}    
\end{subequations}
Let us remark that, when designing the pillar Hamiltonian $H_{G\updownarrow}$ through Eq.~(\ref{eqn:pillar-Ham-invariance}), we only checked for invariance under gauge transformations but not under group automorphisms. 
Correspondingly, finite values of $\Re[d_j]$ and $\Im[d_j]$ violate certain automorphisms in $\textrm{Aut}(Q_8)$.
Therefore, inducing the energy splitting by $H_{G \updownarrow}$ needs to be applied with care when automorphism-modified symmetries are present, to ensure that the elements of $\mathsf{SG}$ \mbox{are not broken}.

\appsectitle{Time-reversal symmetry of the irrep sectors}

In the Methods section of the main text, we demonstrated how a spinful time-reversal operator arises for the Hamiltonian block transforming in the irrep $K = E$ of the gauge group $G = Q_8$.
This spinfulness occurs despite the `microscopic' time-reversal operator $T$, which acts at the level of the constituent orbitals in Fig.~1 of the main text, being manifestly spinless. 
In the present section, we discuss the case of a general gauge group $G$ and irrep $K$, and how the character of the time-reversal symmetry $T^K$ of Hamiltonian block $H^K$ is captured by the Frobenius-Schur indicator in Eq.~(18).

Our discussion is structured as follows. 
First, attempting to generalize Eq.~(15), we compute the matrix element of $U^\dagger U^* \mathcal{K}$ in the irrep basis for a general choice of a gauge group $G$. 
Second, we investigate components of $U^\dagger U^* \mathcal{K}$ corresponding to various choices of irrep $K$.
In particular, we distinguish two cases, depending on whether the irrep $K$ is complex (Case A, no time-reversal symmetry acts on the Hamiltonian block) or self-conjugate (Case B, an effective time-reversal operator can be constructed). Third, we determine whether an effective time-reversal operator, when it exists, squares to $+1$ (spinful TRS) or to $-1$ (spinless TRS). 
We connect the identified three cases to the Frobenius-Schur indicator, and we highlight the special role of $Q_8$.

\bigskip

In general, the Peter-Weyl decomposition always leads us to consider the object $U^\dagger U^* \mathcal{K}$ [left side of Eq.~(15)]. 
By substituting the general expression in Eq.~(11), we find
\begin{eqnarray}
&\phantom{=}& (U^\dagger U^*)_{(K,\lambda,\nu),(K',\lambda',\nu')} \nonumber \\ 
&=& \sum_{g\in G} \! U^*_{g,(K,\lambda,\nu)} U^*_{g,(K',\lambda',\nu')} \nonumber \\
&=& \frac{\sqrt{d_K d_{K'}}}{|G|} \!\sum_{g\in G} \!D_{\nu\lambda}^K(g) D_{\nu'\lambda'}^{K'}(g) = \cdots. \label{eqn:Udag-Ustar}
\end{eqnarray} 
We are specifically interested in whether the microscopic time-reversal operator $T$ mixes various copies of irrep $K$, i.e., when non-trivial components arise for $K = K'$; in such cases, one can find diagonal blocks $\mathcal{T}^K$ through multiplication with a unitary $M$ while keeping the Hamiltonian invariant, as exemplified by Eq.~(17) for the case of the quaternion group. 
In contrast, if time reversal relates different irreps $K \neq K'$, then the construction in Eq.~(17) mixes different Hamiltonians $H^K \neq H^{K'}$, thus invalidating the construction~of~$\mathcal{T}^K$.

Observe that the right-hand side of Eq.~(\ref{eqn:Udag-Ustar}) differs from the standard Schur orthogonality relation by the absence of complex conjugation on the second factor behind the sum. 
This is rectified by replacing the irrep $K'$ by its dual (i.e., complex-conjugate) irrep $K'^*$; then,
\begin{equation}
\cdots \! = \!\frac{\sqrt{d_K d_{K'}}}{|G|} \! \sum_{g\in G} \! D_{\nu\lambda}^K(g) \! \left[D_{\nu'\lambda'}^{K'^*}(g)\right]^*\! \! \! = \! \delta_{K,K'^*} \delta_{\nu \nu'} \delta_{\lambda \lambda'}.
\end{equation}
To proceed, we need to distinguish two cases.

\bigskip

In Case~A, $K \nsim K^*$, i.e., the representations $K$ and $K^*$ are not unitarily equivalent. 
Such a representation is described as complex.
In this case, the components 
\begin{equation*}
(U^\dagger U^*)_{(K,\lambda,\nu),(K,\lambda',\nu')}  
\end{equation*} 
that relate copies of the same irrep $K$ contain the Kronecker symbol $\delta_{K,K^*} = 0$. 
It follows that $U^\dagger T U$ exchanges two different Hamiltonian sectors $H^K \neq H^{K^*}$; therefore, no time-reversal symmetry acts on the Hamiltonian $H^K$.

The `Case A' does not arise for irreps of $Q_8$; however, it generally arises for irreps of the cyclic groups $Z_n$ with $n\geq 3$. 
We briefly discuss this example due to the physical intuition associated with interpreting $Z_n < U(1)$ as the quantization of magnetic fluxes to integer multiples~of~$2\pi/n$. 
Specifically, understanding the elements of the cyclic group as the complex numbers $e^{2\pi i k/n}$ with $k\in \{0,1,\ldots,n-1\}$, one finds $n$ one-dimensional irreps $D^\omega$ indexed by $\omega\in\{0,1,\ldots,n-1\}$, which are defined by $D^\omega(e^{2\pi i k/n}) = e^{2\pi i \omega k/n}$.
Crucially, $[D^\omega]^* = D^{n-\omega}$, meaning that $D^\omega$ is a complex representation whenever $\omega \notin \{0,n/2\}$.

Physically, connections valued in $Z_n$ arise when quantizing magnetic [i.e., $U(1)$] flux to integer multiples of $2\pi\omega/n$. 
Time-reversal operator relates the Hamiltonian block $H^\omega$ to the Hamiltonian block $H^{n-\omega}$, in which the magnetic fields $B$ (fluxes $\Phi$) are replaced by fields $-B$ (fluxes $-\Phi$). 
Time-reversal symmetry is absent within an individual block when $\omega \neq n-\omega$~(mod~$n$). Such Hamiltonians arise as blocks of a CSL with gauge group~$G = Z_n$.

\bigskip

In Case~B, $K \sim K^*$, i.e., the representation $K$ is unitarily equivalent to $K^*$.
Such a representation is described as self-conjugate (and we show below that they can be further specified as being either real or symplectic). 
All irreps of $Q_8$ belong to this category.
In this case, the Kronecker symbol $\delta_{K,K^*}=1$ implies that $U^\dagger U^*$ will have non-trivial blocks relating copies of the same irrep $K$; in contrast, $\delta_{K,K'^*} = 0$ whenever $K \neq K'$.

The unitary equivalence of $K$ and $K^*$ implies the existence of a unitary matrix $C^K$ such~that
\begin{equation}
\label{eqn:self-conjugate-irrep}
D^{K^*}(g) = \big(C^K\big)^\dagger D^K(g) C^K
\end{equation}
for all $g \in G$.
In components, this reads
\begin{equation}
D^{K^*}_{\nu'\lambda'}(g) = \big(C^K\big)^*_{\alpha\nu'} D^K_{\alpha \beta}(g) \big(C^K\big)_{\beta \lambda'}.  
\end{equation}
Substituting this expression in the right-hand side of Eq.~(\ref{eqn:Udag-Ustar}) gives
\begin{eqnarray}
&\phantom{=}&
(U^\dagger U^*)_{(K,\lambda,\nu),(K,\lambda',\nu')} \nonumber \\
&=& \frac{d_K}{|G|}\sum_{g\in G} D^K_{\nu \lambda}(g) \big(C^K\big)_{\alpha \nu'} \big[D^K(g)\big]^*_{\alpha\beta} \big(C^K\big)^*_{\beta \lambda'} \nonumber \\
&=& \big(C^K\big)_{\alpha \nu'} \big(C^K\big)^*_{\beta \lambda'} \frac{d_K}{|G|}\sum_{g\in G} D^K_{\nu \lambda}(g) \big[D^K(g)\big]^*_{\alpha\beta} \nonumber \\
&=& \big(C^K\big)_{\alpha \nu'} \big(C^K\big)^*_{\beta \lambda'} \delta_{\nu \alpha} \delta_{\lambda \beta} \nonumber \\
&=& \big(C^K\big)^*_{\lambda \lambda'} \big(C^K\big)_{\nu \nu'} , 
\end{eqnarray}
where in going from the third line to the fourth line we employed the Schur orthogonality relation, and in the last line we chose the ordering of the components compatible with tensor product.
We find that the block of $U^\dagger U^*$ which relates the $d_K$ copies of $K$ can be expressed compactly~as 
\begin{equation}
\left. U^\dagger U^* \right|_K = \left(C^K\right)^* \otimes C^K.   
\end{equation}
When reading such tensor products, our convention is that the first factors acts on the $d_K$ copies of irrep $K$ (i.e., on index $\lambda$) while the second factor acts on the $d_K$ states transforming within the individual copies of the irrep (i.e., on index $\nu$).

In the spirit of Eq.~(17), we seek a matrix $M^K$ such that $M^K \big(\left.U^\dagger U^*\right|_K\big) \mathcal{K} = \mathscr{T}^K$ is diagonal in the irrep basis (i.e, in the first factor of the tensor product); that is, we require
\begin{equation}
\mathscr{T}^K = \mathbb{1} \otimes \mathcal{T}^K.    
\end{equation}
In order to commute with the Hamiltonian, the matrix $M^K$ can mix the $d_K$ copies of irrep $K$ (first factor in the tensor product) but must act trivially on the $d_K$ orbitals spanning any copy of the representation $K$ (second factor in the tensor product).
Up to an unimportant ambiguity of the overall phase, we find the solution $M^K = (C^K)^\top \otimes \mathbb{1}$. 
With this choice, we~obtain 
\begin{equation}
\mathscr{T}^K \!=\! \left[\big(C^K\big)^\top \otimes \mathbb{1}\right] \cdot \big[\big(C^K\big)^* \otimes C^K\big] \, \mathcal{K} = \big( \mathbb{1} \otimes C^K \big) \, \mathcal{K},
\end{equation}
where in the first factor we utilized the unitarity of $C^K$.
The identity $\mathbb{1}$ appearing as the first factor ensures that every copy of $H^K$ is assigned the same time-reversal operator $\mathcal{T}^K$.

We next inspect whether the effective time-reversal symmetry $\mathcal{T}^K$ acting on the Hamiltonian block $H^K$ is spinless or spinful. 
This is determined by the value of 
\begin{equation}
\label{eqn:general-TRS-square}
\big(\mathcal{T}^K\big)^2 = \big( C^K \mathcal{K} \big)^2 = C^K \big(C^K\big)^*.
\end{equation}
That $C^K (C^K)^* = \pm \mathbb{1}$ generally follows from substituting Eq.~(\ref{eqn:self-conjugate-irrep}) `into itself' as follows.
First, perform complex conjugation of this equation, i.e., 
\begin{equation}
\label{eqn:self-conjugate-irrep-cc}
D^K(g) = \big(C^K\big)^\top D^{K^*}(g)\big(C^K\big)^*.
\end{equation}
Substituting Eq.~(\ref{eqn:self-conjugate-irrep}) on the right-hand side of Eq.~(\ref{eqn:self-conjugate-irrep-cc}) gives
\begin{equation}
 D^K(g) = \big(C^K\big)^\top \big(C^K\big)^\dagger D^K(g) C^K \big(C^K\big)^*,   
\end{equation}
which can be rewritten as the commutation relation
\begin{equation}
\label{eqn:commute-DK-CC}
\Big[D^K(g),C^K \big(C^K\big)^*\Big] = 0.    
\end{equation}
Since $K$ is an irreducible representation and Eq.~(\ref{eqn:commute-DK-CC}) applies for all $g\in G$, Schur's lemma ensures that 
\begin{equation}
\label{eqn:CK-CK-alpha}
C^K \big(C^K\big)^* = \alpha_K \, \mathbb{1}    
\end{equation} 
is a diagonal matrix. 
To reveal the discrete character of the scalar coefficient $\alpha_K$, we first take the transpose of Eq.~(\ref{eqn:CK-CK-alpha}),~i.e., 
\begin{equation}
\big(C^K\big)^\dagger \big(C^K\big)^\top = \alpha_K \, \mathbb{1},
\end{equation} 
which we further multiply from the left by $C^K$ and from the right by $(C^K)^*$, resulting in 
\begin{equation}
\underbrace{C^K \big(C^K\big)^\dagger}_{\mathbb{1}} \underbrace{\big(C^K\big)^\top \big(C^K\big)^*}_{\mathbb{1}} = \alpha_K C^K \big(C^K\big)^*.
\label{eqn:CK-CK-alpha-2}
\end{equation}
Substituting Eq.~(\ref{eqn:CK-CK-alpha}) on the right-hand side of Eq.~(\ref{eqn:CK-CK-alpha-2}) results in 
\begin{equation}
\mathbb{1} = \alpha^2_K \mathbb{1}.    
\end{equation} 
It follows that $\alpha_K = \pm 1$.
Irrep $K$ with $\alpha_K = +1$ (with $\alpha_K = -1$) is described as real (as symplectic).
Returning back to Eq.~(\ref{eqn:general-TRS-square}), we recover that $\alpha_K = +1$ ($\alpha_K = -1$) corresponds to the appearance of spinless (spinful) time-reversal symmetry in the Hamiltonian block~$H^K$.

\bigskip

What remains to be discussed is a practical test that efficiently captures whether a given representation $K$ falls into (1)~Case~A (no time reversal), (2)~Case~B with $C^K (C^K)^* = +\mathbb{1}$ (spinless time reversal), or (3)~Case~B with $C^K (C^K)^* = -1$ (spinful time reversal). 
Further analysis can be used to relate this trichotomy to the Frobenius-Schur indicator, specified in Eq.~(18) of the Methods section.  
We do not reproduce the derivation of the Frobenius-Schur indicator here, but refer instead to the derivation and discussion in Sec.~13.2 of the book by Serre, see Ref.~\citenum{Serre:1977}. 
One general implication that can be shown is that time-reversal symmetry (if present) of $H^K$ for an odd-dimensional representation $K$ is ensured to be spinless.
On the other hand, even-dimensional representations can give rise to either spinless or spinful time-reversal operator, i.e., the spin-$\tfrac{1}{2}$ \& spin-orbit-coupling interpretation of two-dimensional irrep follows only if the Frobenius-Schur indicator evaluates to $\iota_K = -1$.

\bigskip

We point out that the quaternion group $Q_8$ is the smallest group supporting a symplectic representation, i.e., it furnishes the elemental case of CSLs supporting spin-$\tfrac{1}{2}$ (i.e., two-component spinor) Hamiltonian blocks. 
To recognize this special aspect of $Q_8$, note that most of the other groups of order $|G| \leq 8$ are Abelian (therefore supporting only one-dimensional irreps whose time-reversal operator, if present, is spinless). 
The only other non-Abelian groups within this range are $S_3$ (the permutation group of three elements; equivalently, the symmetry group of a triangle: $|S_3| = 6$) and $D_4$ (the dihedral group; equivalently, the symmetry group of a square: $|D_4| = 8$). 
Both $S_3$ and $D_4$ support one two-dimensional irrep; however, this irrep is characterized by $\iota_K = +1$ and, therefore, results in spinless time-reversal symmetry.
This distinct phenomenology is related to the fact that $Q_8$ is a subgroup of $SU(2)$; in contrast, $S_3$ and $D_4$, being symmetry groups of geometric objects, are subgroups of $O(2)$ but not of $SU(2)$.

\appsectitle{Symmetry of the triangular ladder model in Fig.~2\bfsf{b}}
\label{App:Trianguar-Ladder-B-symmetry}

In the Methods section, we presented the Hamiltonian of the triangular ladder model of Fig.~2\bfsf{b} of the main text [Eq.~(21)], and listed expressions for its time-reversal symmetry $\mathcal{T}$ and chiral symmetry $\mathcal{S}$ [below Eq.~(21)], mirror symmetry $\mathsf{M}_x$ [Eq.~(22); present only if we set $t_1 = t_2$], and translation symmetry $\mathsf{T}_x$ [Eq.~(23)]. 
Here, we continue the discussion of the model by explaining how the symmetry of the model enforces the spectral features illustrated in Fig.~\ref{fig:ladders-bands}\bfsf{b} (for $t_1 \neq t_2$) and Fig.~\ref{fig:ladders-bands}\bfsf{c} (for $t_1 = t_2$) of the main text.

We have formerly concluded that $\mathsf{T}_x$ enforced the bands to appear in translation doublets, connected together across the Brillouin zone boundary, which at momentum $k$ exhibit translation eigenvalues $(e^{ik/2},-e^{ik/2})$. 
Correspondingly, we color the energy bands in Supplementary Fig.~\ref{fig:ladders-bands}\bfsf{b} according to the complex phase of their translation eigenvalue. 
Spinful time-reversal symmetry enforces the presence of Kramers pairs at TRIMs.
The commutator $[T_x,\mathcal{T}]=0$ at each TRIM ensures that Kramers pairs carry translation eigenvalues $(\lambda,\lambda^*)$: either $(1,1)$ or $(-1,-1)$ at $k=0$, and $(i,-i)$ at $k=\pi$.
Since the translation eigenvalues of a Kramers pair at $k=0$ are not compatible with a translation doublet, it follows~\cite{Bzdusek:2016} that time-reversal symmetry ties two translation doublets together, forming a butterfly-like spectrum of four connected energy bands as visible in Supplementary Fig.~\ref{fig:ladders-bands}\bfsf{b}.
Due to chiral symmetry, one such quadruplet of bands arises at positive and another one at negative energies.

We conclude with discussing why a fourfold degeneracy is exhibited by the model at energy $\varepsilon=0$ and momentum $k=\pi$ when $t_1 = t_2$ (i.e., the additional $\mathsf{M}_x$ symmetry) is imposed. 
We show the band structure for this choice of parameters (with the bands colored according to their $\mathsf{T}_x$ eigenvalue) in Supplementary Fig.~\ref{fig:ladders-bands}\bfsf{c}.
We summarize in broad strokes how the band closing follows from the algebra of the operators. 
First, the relations $\{\mathcal{S}, \mathcal{H}^E_{2\times}\}=0$, $[\mathcal{T},\mathcal{H}^E_{2\times}]=0$, and $\{\mathcal{S},\mathcal{T}\}=0$ imply that there is a chiral basis in which $\mathcal{S}\propto \sigma_z$, $\mathcal{T} = i \sigma_y \mathcal{K}$, and the Hamiltonian acquires the block-off diagonal form
\begin{equation}
W\mathcal{H}^E_{2\times}W^\dagger(k)=\begin{bmatrix} 0 & D(k) \\ D^\dagger(k) & 0 \end{bmatrix}.
\end{equation}
The commutation relations $[\mathsf{M}_x,\mathcal{S}]=0$ and $[\mathsf{M}_x,\mathcal{T}]=0$ further imply that one can additionally diagonalize $M_x$ while staying within a chiral basis.
For the model in Eq.~(21), we find that such diagonalization is achieved with the matrix
\begin{equation}
W =
\begin{bmatrix}
 \frac{1}{\sqrt{2}} & 0 & 0 & 0 & 0 & \frac{1}{\sqrt{2}} & 0 & 0 \\
 \frac{1}{\sqrt{2}} & 0 & 0 & 0 & 0 & -\frac{1}{\sqrt{2}} & 0 & 0 \\
 0 & 0 & 0 & i & 0 & 0 & 0 & 0 \\
 0 & 0 & 0 & 0 & 0 & 0 & -1 & 0 \\
 0 & \frac{1}{\sqrt{2}} & 0 & 0 & -\frac{1}{\sqrt{2}} & 0 & 0 & 0 \\
 0 & \frac{1}{\sqrt{2}} & 0 & 0 & \frac{1}{\sqrt{2}} & 0 & 0 & 0 \\
 0 & 0 & i & 0 & 0 & 0 & 0 & 0 \\
 0 & 0 & 0 & 0 & 0 & 0 & 0 & -1 \\
\end{bmatrix}.
\end{equation}
Importantly, if we narrow attention to momentum $k=\pi$, we find that mirror eigenvalues of the positive-chirality block (i.e., the first four elements of $W \mathsf{M}_x W^\dagger$) are $(-i,+i,+i,+i)$ while the negative-chirality block carries mirror eigenvalues $(+i,-i,-i,-i)$.
Since $k=\pi$ is a mirror-invariant momentum, only states with the same mirror eigenvalue can be coupled by the off-diagonal block $D(\pi)$ of the Hamiltonian. 
Owing to the imbalance of the mirror eigenvalue content of the two chiral sectors, $D(\pi)$ itself takes the off-diagonal form
\begin{equation}
\label{eqn:off-diagonals}
D(\pi) = \begin{bmatrix} 0 & d \\ d^\dagger & \mathbb{0} \end{bmatrix}   
\end{equation}
where $d$ is a $1\times 3$ array. 
The rectangular shape of the blocks in Eq.~(\ref{eqn:off-diagonals}) ensures, by virtue of the rank-nullity theorem, that $\det [D(\pi)] = 0$, meaning that the Hamiltonian has zero-energy states at $k=\pi$. 
In combination with the butterfly-like spectrum and sublattice symmetry present for general $t_{1,2}$, it follows that the zero-energy states for $t_1 = t_2$ correspond to a four-fold degeneracy.

\appsectitle{Partial polarization}
\label{App:partial-polarization}

We elaborate on the definition and computation of the partial polarization. 
For the model in Fig.~2\bfsf{a} of the main text, we have discussed that for $t_1 \neq t_2$ the Hamiltonian in the $E$ sector possesses both time-reversal $\mathcal{T} = \sigma_0 \otimes i \sigma_y \mathcal{K}$ and rotation $\mathsf{C}_{2z} = \sigma_x \otimes i (\sigma   _x + \sigma_y) / \sqrt{2}$ symmetries, both of which square to $ \mathcal{T}^2 = -1 = \mathsf{C}_{2z}^2$ and commute $\left[\mathcal{T}, \mathsf{C}_{2z}\right] = 0$. Note that in a one-dimensional system, this rotation symmetry effectively acts like a mirror symmetry. 

Kramers' theorem dictates that each eigenstate at momentum $k$ comes with a time-reversed partner at $-k$ at the same energy. 
This allows us to separate each doublet of bands into time-reversal channels. 
The partial polarization $P^s$ is the polarization of one such channel~\cite{Kane_2006}. 
The total polarization of the doublet is the sum of the partial polarizations originating from the time-reversal channels $P = P^1+P^2$. 
One can show that the presence of spinful time-reversal symmetry leads to $P^1 = P^2$, meaning that each time-reversal channel assumes the same polarization. 
Then, making use of $\mathsf{C}_{2z}$, one can show that $P^s = -P^s \,\, \mathrm{mod}~1$ ($s=1,2$), which quantizes the partial polarization values to $0$ or $1/2$.~\cite{Ortix_2016}
It follows that the (total) polarization of a doublet always vanishes, even when the partial polarization is nontrivial.

In one-dimensional insulators, the polarization $P$ and the Berry phase $\gamma$ are related to each other as $P = \gamma / 2 \pi$. 
As a result, the computation of the partial polarization for the systems shown in Fig.~2\bfsf{a},\bfsf{b} of the main text [Hamiltonians in Eqs.~(19) and (20)] consists of the calculation of the Berry-phase for one of the time-reversal channels. 
To consider only one of the time-reversal channels, one needs to choose the states $\ket{u^s(k)}$ in such a way that the projector to the channel $\mathsf{P}(k) = \ket{u^s(k)} \bra{u^s(k)}$ is a continuous function of $k$. Since the channels in each doublet are degenerate only at $k=0$ and $\pi$, this can be achieved. 
For example, for the two time-reversal-related bands displayed in black (or in green) at energy $\varepsilon >0$ in Fig.~2\bfsf{c} of the main text, one channel corresponds to the band with a positive derivative at $k=0$ and a negative derivative at $k=\pi$ (while the other channel corresponds to the band with opposite signs of the derivatives).
The non-zero value of the Berry phase indicates the presence of Kramers-degenerate zero modes on each end of the wire.

\end{bibunit}
\let\oldaddcontentsline\addcontentsline     
\renewcommand{\addcontentsline}[3]{}        
\bibliography{ref}
\let\addcontentsline\oldaddcontentsline